\documentclass[a4paper,fleqn,usenatbib]{mnras}

\usepackage[T1]{fontenc}
\usepackage{ae,aecompl}
\usepackage{graphicx}	
\usepackage{amsmath}	
\usepackage{amssymb}	
\usepackage{dblfloatfix}
\usepackage{color,soul}
\usepackage{indentfirst}
\usepackage{url}
\usepackage{scalerel}
\usepackage{placeins}
\usepackage{float}
\restylefloat{table}
\usepackage{lpic}
\usepackage{siunitx}
\usepackage{multirow}
\usepackage{caption}
\usepackage{subcaption}
\usepackage[textwidth=1.5cm,shadow]{todonotes}	
\usepackage[normalem]{ulem}    
\usepackage{color,soul}
\usepackage{wrapfig}	
\usepackage{newtxtext,newtxmath}
\usepackage{makecell}

\newcommand{\orcid}[1]{\href{#1}{\includegraphics[scale=0.04]{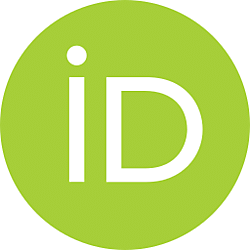}}}

\newcommand{\lagn}{L_{\rm AGN}}

\newcommand{\msun}{{\rm M_\odot}}

\newcommand{\mach}{\mathcal{M}}
\newcommand{\Ward}{\textit{W24} }

\usepackage{lineno}

\title[Brighter AGN produce denser outflows]{Tracing AGN Feedback Power with Cool/Warm Outflow Densities: Predictions and Observational Implications}

\author[Almeida et al.]{
Ivan Almeida$^{1, 2}$\thanks{E-mail: ivalmeida@mpia.de}\orcid{https://orcid.org/0000-0001-6018-2852},
Tiago Costa$^{2}\orcid{https://orcid.org/0000-0002-6748-2900}$,
Chris M. Harrison$^2$\orcid{https://orcid.org/0000-0001-8618-4223} \&
Samuel R. Ward$^{2,3}$\orcid{https://orcid.org/0000-0001-5345-0900}
\\
$^1$Max Planck Institute for Astronomy, Königstuhl 17, D-69117 Heidelberg, Germany \\
$^2$School of Mathematics, Statistics and Physics, Newcastle University, NE1 7RU, UK\\
$^3$Center for Computational Astrophysics, Flatiron Institute, New York, NY 10010, USA
}

\date{Accepted 2026 January 27. Received 2026 January 27; in original form 2025 August 25}
\pubyear{2025}

\begin{document}
\label{firstpage}
\pagerange{\pageref{firstpage}--\pageref{lastpage}}
\maketitle

\begin{abstract}

Winds launched at the scale of the accretion disc or dusty torus in Active Galactic Nuclei (AGN) are thought to drive energy-conserving outflows that shape galaxy evolution.
The key signature of such outflows, the presence of a hot ($T \gtrsim 10^9 \, \rm K$), shocked wind component, is hard to detect directly. Observations of AGN outflows typically probe a separate outflow phase: cool/warm gas with $T \lesssim 10^5 \, \rm K$.
Here, we show that the density of cool outflowing gas scales with AGN luminosity, serving as an indirect diagnostic of the elusive hot, shocked wind.
We use hydrodynamic simulations with the moving-mesh code \texttt{AREPO} to target the interaction between a small-scale AGN wind of speed $\approx 10^4 \, \rm km \, s^{-1}$ and galactic discs containing an idealised, clumpy interstellar medium (ISM). Through a new refinement scheme targeting rapidly-cooling, fast-moving gas, our simulations reach a resolution of $\lesssim 0.1 \, \rm pc$ in the cool, outflowing phase.
We extract an ensemble of cool clouds from the AGN-driven outflows produced in our simulations, finding that their densities increase systematically with AGN wind power and AGN luminosity. Moreover, the mass distribution and internal properties of these cloudlets appear to be insensitive to the initial properties of the ISM, and shaped mainly by the dynamics of radiative, turbulent mixing layers.
The increase in cool outflow density with kinetic wind power and AGN luminosity has profound implications for observational estimates of outflow rates and their scaling with AGN luminosity. Depending on the available outflow and density tracers, observationally-derived outflow rates may be overestimated by orders of magnitude.
\end{abstract}




\section{Introduction}\label{sec:introduction}

Evidence suggests that supermassive black holes (SMBH) inhabit the centres of all massive galaxies \citep{Kormendy2013}. In the local Universe, they have typical masses of $10^6-10^{10}\msun$, amounting to $\gtrsim 0.1\%$ of the galactic bulge mass \citep{Ferrarese2000, McConnell2011, vandenBosch2016}, concentrated within a small region with radius $\approx 2$ AU $(M_{\rm BH} / 10^8 \msun)$. To grow to their current masses, these SMBHs must have accreted from their surrounding gas reservoir, ``lighting up'' as active galactic nuclei (AGN), emitting across the full electromagnetic spectrum \citep{Marconi2004, EHTC2019, DeMenezes2020}.

AGN release vast amounts of energy into their host galaxies, producing luminosities up to $\gtrsim 10^{48}$ erg s$^{-1}$ \citep[][]{Bischetti2017, Onken2020}, and launching powerful wide-angle winds and collimated jets \citep{Fabian2012, Abramowicz2013, King2015, Saikia2022}. If this energy couples efficiently to ambient gas, it can profoundly affect the evolution of the host galaxies through an interaction known as ``AGN feedback''. 
State-of-the-art cosmological simulations of galaxy evolution rely on strong AGN feedback to reproduce the observed properties of massive galaxies, including the high-end of the stellar mass function, galaxy colours, morphologies, sizes and star formation properties \citep{Hirschmann2014, Vogelsberger2014, Schaye2015, Dubois2016, Henden2018, Pillepich2018, Dave2019, Dubois2021, Ni2022, Dolag2025}. 
However, due to resolution limitations and the exclusion of key physical processes (e.g. general-relativistic-radiation-magnetohydrodynamics), such large-scale cosmological simulations follow AGN feedback using heuristic ``subgrid models'' that do not explicitly track the impact of jets, radiation and winds on the interstellar medium (ISM).

Recognised among the most efficient of AGN feedback processes is the interaction of fast ($v \gtrsim 0.1c $) winds launched from accretion disc scales \citep{Tombesi2012, Laha2021, Matzeu2023, Xu2025} with the ISM and the circumgalactic medium \citep[e.g.][]{King2003, Zubovas2012, Faucher2012, Wagner2013, Costa2014, Costa2020}. These winds are expected to decelerate through strong shocks, forming large-scale, energy-conserving bubbles that are extremely hot ($T \sim 10^9 \-- 10^{10} \, \rm K$), and tenuous ($n_{\rm H} \lesssim 10^{-2} \, \rm cm^{-3}$). These bubbles push into ambient gas generating large-scale galactic outflows that can strongly impact galaxy evolution \citep[see e.g.][]{Costa2014}.

In classic, spherical outflow solutions \citep[e.g.][]{King2003, King2005}, the post-shock temperature of outflowing gas is typically $\gtrsim 10^7 \, \rm K$. However, cool gas is expected to condense out of the shocked ambient gas when its characteristic outflow timescale exceeds the cooling timescale \citep{Zubovas2014, Costa2014, Richings2018}. Since they form directly out of an already rapidly outflowing, mass-loaded gas phase, cool ``shells'' are able to retain the high momentum flux and kinetic power of the hot outflow component \citep[e.g.][]{Costa2014, Ward2024, Zubovas2024}. In classic spherical outflow solutions, hot- and cool outflow phases are strongly coupled.

However, the interaction between a fast wind with a more complex density field can dramatically modify the mass, momentum and kinetic energy content of the outflowing cool phase. 
\citet{Wagner2013} simulate the impact of winds with a two-phase, clumpy ISM. Their simulations predict that the energy-driven bubble expands through low density channels, coupling to cool clouds through its ram pressure mainly \citep{Nayakshin2014}, in contrast to spherical outflow solutions where thermal pressure plays a dominant role.
\cite{Bourne2014} directly compare the impact of `energy-conserving' bubbles on homogeneous and clumpy media, finding strong decoupling between a purely hot, outflowing phase and cooler, dense ISM material, which is accelerated less efficiently in inhomoegenous media. 
\citet{Torrey2020} characterise the multi-phase structure of outflowing gas in their simulations of AGN winds interacting with galaxies with a resolved multi-phase ISM, noting the formation of large cavities filled with hot, low-density gas in the central galactic regions. They find no prominent cool component in gas with radial velocity $\gtrsim 200 \, \rm km \, s^{-1}$, with faster outflows dominated by a hot component with $T \gtrsim 10^6 \, \rm K$. Using simulations reaching sub-pc resolution, \citet{Ward2024} directly compare the multi-phase properties between outflows propagating through smooth, homogeneous media and those expanding through clumpy media.
Instead of precipitating from shocked ambient gas, cool outflowing gas forms within turbulent mixing layers comprising AGN wind and pre-existing ISM material, as envisaged in \citet{Gronke2018, Schneider2018, Ji2019, Fielding2020, Sparre2020, Dutta2025}. 
In \citet{Ward2024}, these mixing layers cool radiatively on timescales $\ll \, \rm Myr$, much shorter than the outflow timescale, giving rise to a population of cool ($T \sim 10^4 \, \rm K$) clouds embedded in the hot outflow, with speeds up to $\approx 500 \, \rm km \, s^{-1}$.  \citet{Ward2024} report inefficient coupling between the hot, energy-conserving bubble and the cool phase, finding much lower momentum flux and kinetic powers in the latter.
The recent hydrodynamic simulations modelling the interaction between AGN winds and turbulent gas media presented in \cite{Zubovas2024}
find similarly low momentum and kinetic power coupling efficiencies for cool gas.

Hot outflow phases are predicted to dominate the energy budget of AGN-driven outflows \citep{Costa2015, Torrey2020, Ward2024, Zubovas2024}, acting as the ``driving engine'' of the large-scale outflow. 
In energy-driven outflows, there are in reality two ``hot phases'': (i) a shocked ambient gas component with $T \gtrsim 10^7 \, \rm K$ that may be detectable through spatially extended X-ray or radio emission \citep[e.g.][]{Costa2014a, Nims2015}, and (ii) the energy-conserving, shocked wind bubble. 
While extended X-ray emission has been detected in a few cases \citep{Feruglio2013, Greene2014}, observational probes \citep[e.g.][]{Harrison2018} of AGN-driven outflows typically trace ionised gas phases produced in warm and cool gas with $T \lesssim 10^6 \, \rm K$ through emission lines such as H$\alpha$ or [OIII] \citep{Carniani2016, Zakamska2016, Marasco2023, Riffel2023, Musiimenta2024, Vayner2024, Liu2025}, and molecular lines such as CO and OH \citep{Cicone2014, Feruglio2015, Brusa2018, Fluetsch2019, Puglisi2021, Molina2023}. 
A major obstacle to the observational validation of the energy-driven outflow scenario is the exceedingly low expected emissivity of the shocked wind component on large scales \citep{Faucher2012, Nims2015, Costa2020}, which makes a direct detection extremely challenging. 

In this study, we investigate the properties of the \textit{cool outflowing phase}, probing how these sensitive they are to properties of this invisible energy-driven bubble. As shown by \citet{Ward2024}, cool gas with $T \sim 10^4 \, \rm K$ clumps into outflowing cloudlets of size $\lesssim 10 \, \rm pc$ and number density $n_{\rm H} \gtrsim 40$ cm$^{-3}$, which we here refer to as ``cool gas clouds'' (CGCs). 
We use the simulations presented in \cite{Ward2024} --hereafter referred to as \textit{W24}-- and an expanded set of new simulations with enhanced resolution in fast-moving cool gas. 

The structure of this paper is as follows: Section \ref{sec:methods} outlines our suite of simulations, including new methods to efficiently select and resolve the properties of outflowing cool gas clouds. In Section \ref{sec:results}, we describe the properties of the CGCs, investigating their dependence on the initial conditions and AGN luminosity. 
In Section \ref{sec:discussion}, we present a physical interpretation of our findings, highlight some important implications of our predictions for observations of AGN-driven outflows, and discuss future directions/improvements. Finally, in Section \ref{sec:summary} we summarise our main conclusions.

\section{Hydrodynamic Simulations}\label{sec:methods}

\subsection{AGN winds interacting with clumpy discs}\label{subsec:sims-data}

We use the hydrodynamical simulations presented in \textit{W24}. These constitute a series of numerical experiments designed to investigate the interaction between a fast, small-scale AGN accretion disc wind and an initially clumpy ISM. In this paper, we also perform a number of new simulations. These include a new mesh refinement method, which is applied specifically to rapidly-cooling, fast-moving gas \citep[][]{Rey2024}, resolving the formation and destruction of cool cloudlets within AGN outflows in greater detail (see Section \ref{subsec:high-res-sims}). From now on, whenever we use the term `cool gas', we refer to gas with temperatures $\leq 2 \times 10^4 \, \rm K$, adopting the convention followed in theoretical studies of multi-phase outflows \citep[e.g.][]{Costa2015, McCourt2018, Gronke2018, Tan2024}. The gas phase with temperatures greater than $2 \times 10^4 \, \rm K$ is refered to as `hot phase'. We experimented varying this threshold to $10^5$K, finding no changes in our results (see Section \ref{subsec:find-clouds}).

Our simulations are performed with the moving-mesh, hydrodynamic code \texttt{AREPO} \citep{Springel2010, Pakmor2016, Weinberger2020}. Small-scale quasar winds are modelled using the \texttt{BOLA} (BOundary Layer for AGN) framework presented in \cite{Costa2020}.
\texttt{BOLA} consists of a rigid, boundary surface discretised using a HEALPix pixelisation \citep{Gorski2005} into a number (here 256) solid angle elements of equal area.
A boundary surface is formed at the interface between two, thin spherical layers of \texttt{AREPO} cells. Mass, momentum, and energy fluxes are set at the boundary, independently for each solid angle element. The cells within the inner layer are excluded from hydrodynamic calculations, acting as ghost cells. Conversely, the cells in the outer layer are allowed to evolve hydrodynamically. \texttt{BOLA} is positioned at the disc centre, with a radius of $r_{\rm BOLA} = 10$ pc. This scale is smaller than the expected free-expansion radius \citep{Costa2020} and thus allows us to resolve the formation of a strong reverse shock.

The parameters describing the small-scale winds considered in this study closely follow the known properties of ultra-fast outflows \citep{Tombesi2010, Tombesi2012, Chartas2014, Matzeu2023, Xu2025}, whose impact on ambient gas in galaxies is thought to result in strong quasar feedback \citep{King2003, Faucher2012, Costa2014}.
While \texttt{BOLA} provides full anisotropic control of the imposed outflow solution \citep[see][]{Costa2020}, we here consider fully isotropic winds, in line with (e.g., \citealt{Faucher2012}).
We assume a velocity of $v_{\rm wind} = 10^4$ km s$^{-1}$ at injection, except in two simulations where we explore $v_{\rm wind} = 5 \times 10^3$ km s$^{-1}$ and $v_{\rm wind} = 3 \times 10^4$ km s$^{-1}$ (Section \ref{subsec:Impact-Initial-ISM}). We assume an initial wind temperature of $T_{\rm wind} = 10^6$ K. As long as the wind is highly supersonic at injection, the case here, the initial temperature does not have a dynamical impact on the large-scale outflow solution \citep{Costa2020}.
We explore AGN luminosities, $L_{\rm AGN}$, ranging from $10^{43}$ to $10^{47}$ erg s$^{-1}$. Assuming a wind momentum flux of $L_{\rm AGN}/c$ implies that the wind power scales with luminosity as $\dot{E}_{\rm k} \propto L_{\rm AGN}$ and that the kinetic coupling efficiency is $\dot{E}_{\rm k} / L_{\rm AGN}
 \, = \, v_{\rm wind}/2c \, = \, 0.0167 \left( v_{\rm wind} / 10^4 \, \rm km \, s^{-1} \right)$, as described in \citet{Costa2020}.
 Note that the wind prescribed on the surface boundary is smooth, with all the outflowing `cool clouds' discussed later in this paper forming at larger scales through a complex interaction between this wind and the ISM \citep[see][]{Ward2024}. These cool, outflowing cloudlets constitute an emergent property of the large-scale outflows and a full prediction of our simulations.

We assume a primordial of H and He (a hydrogen fraction of 76\%) and associated radiative cooling assuming an optically thin gas, in ionization equilibrium \citep{Katz1996, Weinberger2020}. We exclude a UV background -- for this work, we choose to isolate the effect of AGN winds on the initial clumpy ISM model, excluding any other effect on the ISM state. This can cool gas down to a minimum temperature $\sim 10^4$~K.

We also perform a number of test simulations with metal line cooling \citep{Vogelsberger2014}, for $\lagn = 10^{44}$~erg s$^{-1}$ and $\lagn = 10^{46}$~erg s$^{-1}$. In these simulations, gas contained within cool clumps is assumed to have solar metallicity and the hot background medium is set to have metallicity equal to $10^{-4}$ of the solar metallicity. Using this setup, we probe two different cooling floors: $10^4$~K and $10^2$~K.

In \textit{W24}, various initial ISM configurations are explored. Although \textit{W24} also test the impact of winds on a smooth disc, here we focus on the more realistic clumpy ISM configurations.
For the clumpy case, a two-phase ISM is manually created within a disc spanning a radius of $2$ kpc and a height of $1$ kpc, following \citet{Mukherjee2016, Tanner2022}. The mean total number density in the disc is $\overline{n}_{\rm init} = 5$ cm$^{-3}$ --considering an average molecular weight for ionized gas $\mu = 0.6$--, with an average temperature of $T_{\rm clump} = 10^4$ K.
The total disc gas mass is $M_{\rm disc} = 1.4 \times 10^9 \msun$. \textit{W24} explore varying both $\overline{n}_{\rm init}$ and $M_{\rm disc}$, finding they modify outflow rates by a factor of 2, yet maintaining a similar multi-scale structure across all simulations with an initially clumpy ISM, showing the presence of small cloudlets, and keeping the same outflow shape.

The cool clumps and the surrounding hot background gas are initialised in pressure equilibrium, with the background density and temperature set at $n_{\rm H, background} = 10^{-2}$ cm$^{-3}$ and $T_{\rm background} = 10^7$ K, respectively.
The cool clumps are created using the PyFC python package \citep{Wagner2012}. This generates a random, 3D scalar-field from a given-probability distribution function with a fractal spatial correlation \citep{Lewis2002,Sutherland2007}. This fractal structure of the cool gas follows a log-normal distribution and a Kolmogorov power-law spectrum; further details are available in Section 2.3 of \textit{W24}. The cool phase is parametrized by the average largest clump size $\lambda$ (values showed in Table \ref{tab:simulations-params}) with explored sizes ranging from $40$ pc to $333$ pc. Figure 1 in \textit{W24} illustrates these initial conditions, showing a granular ISM for smaller average clump sizes and fewer, larger clumps for the larger average sizes. This clumpy and disc structure aims to qualitatively capture observed fractal ISM substructures \citep{Wagner2011, Wagner2012b, Mukherjee2016, Tanner2022}.

\subsection{Refining on fast, cool gas}\label{subsec:high-res-sims}

The simulations described in \textit{W24} are performed a target mass resolution of $M_{\rm cell} = 100 \ \msun$, with the exception of \textit{W24}'s high-resolution simulation, which achieves $M_{\rm cell} = 10 \ \msun$. As highlighted in Appendix A of \textit{W24}, increasing the resolution from $100 \ \msun$ to $10 \ \msun$ does not alter some global characteristics of the outflow, such as the mass outflow rate and kinetic luminosity. The properties of the hot phase, in particular, remain unchanged.
However, the outflowing cool gas cloud size was found to decline with increasing resolution, echoing the non-convergence seen in studies of cool gas on galactic scales \citep{vandeVoort2019, Hummels2019}.

Achieving convergence in the cool phase is computationally challenging, requiring extremely high resolution, as noted by \citet{Yirak2010} and \citet{Gronke2018}. For example, the computational time for the high-resolution simulation in \textit{W24} (with $M_{\rm cell}=10 \ \msun$) was $\approx 20$ times longer than for the fiducial simulation (with $M_{\rm cell}=100 \ \msun$).
With cool gas non-convergence and the computational cost of our previous experiments in mind, we here use mesh refinement in a number of new simulations with enhanced resolution only on cool outflowing gas.

\begin{figure*}
    \centering
    \includegraphics[width=\linewidth]{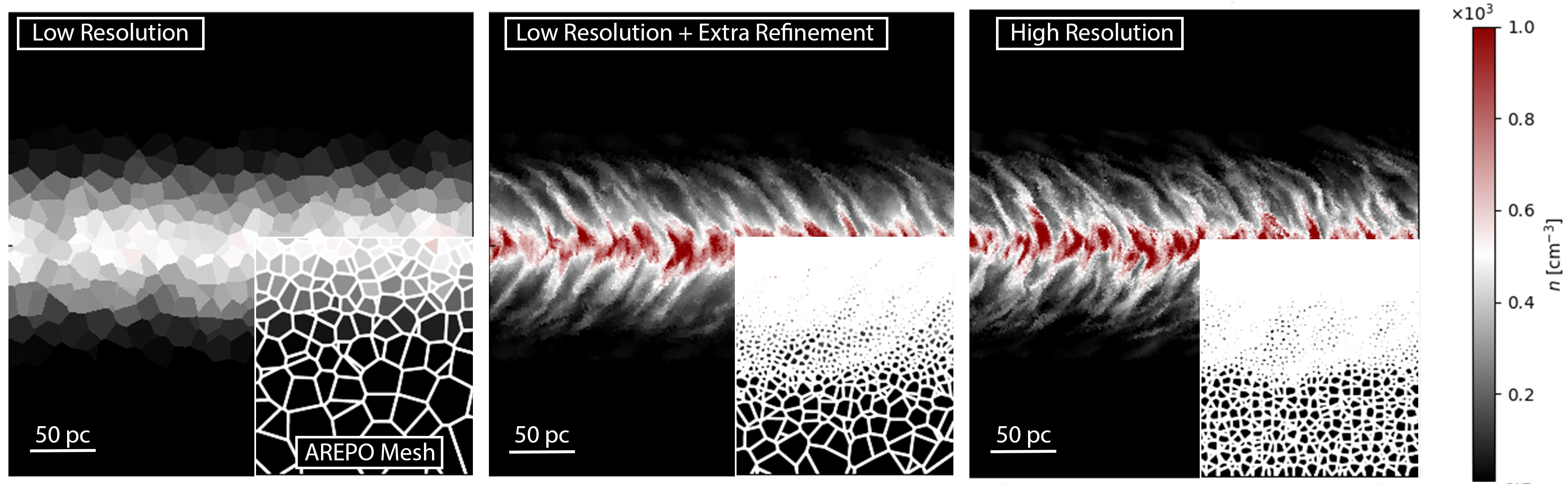}
    \caption{In this test, the initial condition featured a flow with two distinct fluids—a central layer immersed within a surrounding background fluid—each possessing different densities and moving at different speeds (see Section \ref{subsubsec:refinement-tests}). The left panel displays a mass resolution of $10^5 \ \msun$, while the right panel has a mass resolution of $10^2 \ \msun$, 1000 times smaller. The central panel implements our additional refinement technique, with a resolution boost factor of $\beta = 1000$ which enhances the resolution in the regions of interest, for the faster and denser gas. The panels show the gas $20 t_{\rm KH}$ after the start, both the High-Resolution and the new refinement technique simulations exhibit the expected ripples resulting from the Kelvin-Helmholtz instability, which are not captured in the Low-Resolution simulation. Notably, the central panel achieved these results in approximately 37\% of the time required for the right panel, demonstrating the efficiency of the enhanced refinement scheme.}
    \label{fig:refinement-test-KH}
\end{figure*}

\subsubsection{Refinement scheme}
In its standard application, \texttt{AREPO} refines and de-refines cells, so as to maintain $M_{\rm cell}$ within a factor 2 of a target mass $M_{\rm target}$. The goal of our new refinement scheme is to achieve higher resolution in fast-moving cool clouds, maintaining $M_{\text{target}} = 100 \, \text{M}_{\odot}$ elsewhere.

We reduce the target mass by a factor $\beta$ if:

\begin{itemize}
    \item[\textbf{a.}] $M_{\rm cell}$ exceeds a cooling mass, $M_{\rm cool}$ representing the mass scale of the fragmented gas,
    \item[\textbf{b.}] if the local Mach number, $\mach = v/c_\mathrm{s}$, is higher than $\mach_{\rm crit}$ ,
\end{itemize}

\noindent
where $v$ is the gas speed and $c_s$ is the speed of sound.

The first criterion follows \citet{Rey2024}.
We define $M_{\rm cool}$ based on the cooling length $l_{\rm cool} \, = \, c_\mathrm{s} t_{\rm cool} = 3 c_snk_\mathrm{B}T/(2\mu\Lambda)$, where $n$ is the gas number density, $k_\mathrm{B}$ is the Boltzmann constant, $T$ is the temperature, $\mu$ is the mean molecular weight, and $\Lambda$ is the net cooling rate, with units of erg s$^{-1}$ cm$^{-3}$ \citep{Katz1996}.
 We define the cooling mass as:

\begin{equation}
    M_{\rm cool} = \rho l_{\rm cool}^3.
    \label{eq:cooling-mass}
\end{equation}

The first refinement criterion targets regions for which rapid cooling is expected, i.e. $M_{\rm cell} > M_{\rm cool}$. However, this criterion is also satisfied in initial ISM patches that have not yet significantly interacted with the AGN outflow. To further separate the original ISM cool gas from the outflowing cool gas, we chose to refine only if the local Mach number $\mach > \mach_{\rm crit}$.
Experimentally, we have found that setting $\mach_{\rm crit} \, = \, 1$ effectively filters out the low-velocity ISM component, while reproducing a highly-resolved, cool outflowing phase with very similar properties as in the high resolution simulation of \textit{W24}.
Observations and simulations \citep{DalPino2005, Gendron2018, Scholtz2020, Vayner2024} confirm that gas with characteristics similar to our cool outflowing clouds generally moves supersonically, with $\mach > 1$. We also emphasise that more than $\approx 95\%$ of the outflowing gas with $T < 10^6$ K satisfies $\mach > 1$ in the simulations shown in Table \ref{tab:simulations-params}, this number varies slightly depending on $\lagn$, it is $\approx 95\%$ for $\lagn = 10^{43}$ erg s$^{-1}$ and $\approx 100\%$ for $\lagn \geq 10^{46}$ erg s$^{-1}$.

In summary, for cells with $M_{\rm cell} > M_{\rm cool}$ and $\mach \geq 1$, we decrease the target mass by a factor of $\beta$ as
\begin{equation}
    M_{\rm target} = \left\{
  \begin{array}{ c l }
    (100 \beta^{-1}) \ \ \msun & \quad \textrm{if } \mach \geq
    \mach_{\rm crit} \text{ and } M_{\rm cell} > M_{\rm cool}; \\
    100 \ \ \msun                 & \quad \textrm{otherwise}.
  \end{array}
    \right.
  \label{eq:refinement-criteria}
\end{equation}

\noindent The gas cells are refined and de-refined as needed based on this criterion. In order to prevent steep cell size gradients, we trigger additional refinement, keeping volume ratios between adjacent cells below a factor 8.

\begin{table}
\centering
\begin{tabular}{l|ccc}
\hline \hline
\textbf{Model} & LR & Our Method  & HR \\ \hline
\textbf{Target Mass ($\msun$)}$^A$ & $10^5$ & $10^5$ / $10^2 \ ^*$ & $10^2$ \\
\textbf{\textbf{$t_{\rm sim}/t_{\rm KH}$}}$^B$ & $176$ & $12.8$ & $4.7$ \\
\textbf{Speed Ratio}$^C$ & 37.4x & 2.7x & 1x \\ \hline \hline
\end{tabular}
\caption{Comparison of computational speeds across different resolution models --Low-Resolution (LR), Our Method ($\beta = 10^3$), and High-resolution (HR)-- for the KH test performed, as shown in Figures \ref{fig:refinement-test-KH} and \ref{fig:refinement-test-KH-histogram}. This table lists the duration of each model for a fixed amount of computational resources (76 cpu-h). \textbf{A:} Target mass for the numerical cells. \textbf{B:} Total simulated time in $t_{\rm KH}$. \textbf{C:} Speed ratio, normalized to the high-resolution model, indicating relative computational efficiency.
\newline \small{* The base target mass is $10^5 \ \msun$, for the cool outflowing gas, we reduced the target mass to $10^2 \ \msun$.}}
\label{tab:KH-cost-comparison}
\end{table}

\subsubsection{Testing the Refinement Method}\label{subsubsec:refinement-tests}

Our test case includes a dense gas layer in relative motion with respect to a surrounding lower-density medium in pressure equilibrium. This configuration leads to the Kelvin-Helmholtz (KH) instability, which amplifies waves at the interface, rapidly evolving into complex, turbulent, and vortical structures that cause extensive mixing.

The background medium initially has number density $n_{\rm H} = 20$ cm$^{-3}$, temperature $5 \times 10^4$ K, and null velocity. The central layer has a height of $35$ pc, density (temperature) ratio between the central layer and the background is $100$ ($0.01$), with a sinusoidal perturbation amplitude of $8$ pc and a length of $10$ pc. The central denser layer moves at 50 km s$^{-1}$ (Mach number $\mathcal{M} \approx 2$). For this setup, the Kelvin-Helmholtz timescale is $t_{\rm KH} \approx 0.3$ Myr. We perform three different simulations, assuming primordial cooling:

\begin{itemize}
    \item \textbf{Low-Resolution} simulation with a target mass of $10^5 \ \msun$.
    \item \textbf{High-Resolution} simulation with a target mass of $10^2 \ \msun$.
    \item A simulation we name \textbf{Our Method} using our new refinement scheme, with an overall target mass of $10^5 \ \msun$, but a mass resolution boost of $\beta = 10^3$ in rapidly-cooling, supersonic gas, following Equation \eqref{eq:refinement-criteria}.
\end{itemize}

Cooling times are initially long in the central slab, as its temperature remains at the cooling floor of $10^4$ K. Extra-refinement only occurs when there has been significant mixing and the cooling length decreases.
By $t \sim t_{\rm KH}$, the slab temperature rises above $10^4$ K, its cooling rate increases and our refinement criteria are satisfied (see Eq. \eqref{eq:cooling-mass}). For the mixed gas, the cooling timescale is $t_{\rm cool} \lesssim 10^{-3}$ Myr.

Figure \ref{fig:refinement-test-KH} shows the number density in our three test simulations at $20 t_{\rm KH}$, after the instability has fully set in.
The left-hand panel shows results for the Low-Resolution simulation, which shows no clear sign of the instability even at $20 t_{\rm KH}$.
The simulation performed with Our Method (central panel) closely resembles the High-Resolution simulation (right-hand panel), both showing the vortical features absent in the Low-Resolution simulation.

\begin{figure}
    \centering
    \includegraphics[width=\linewidth]{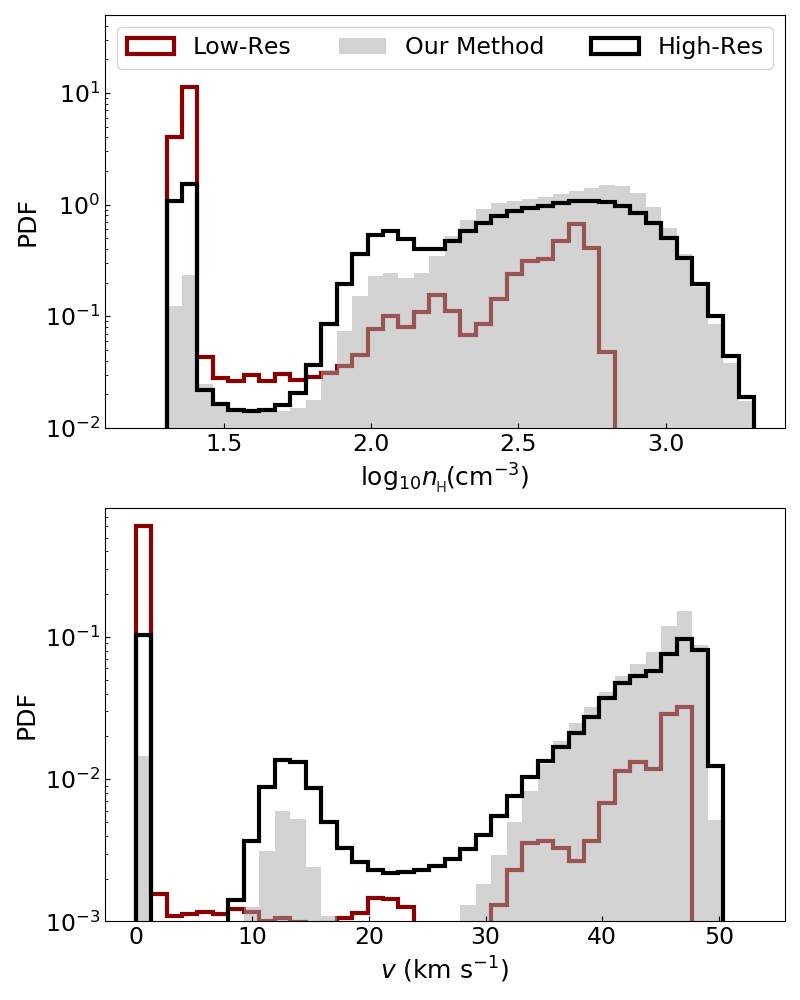}
    \caption{Density (top) and velocity (bottom) distribution for the Low-Resolution (dark red line), High-Resolution (black line) and the intermediate simulation using Our Method (grey area). Our Method presents a similar distribution to the high-resolution simulation, for the fast and dense gas. Overall, the shape of the grey area, resembles the black line (high-resolution) more than the red line (Low-Resolution).}
    \label{fig:refinement-test-KH-histogram}
\end{figure}

The density and velocity distributions of the simulation cells, shown in Figure \ref{fig:refinement-test-KH-histogram}, show that our refinement scheme successfully reproduces the High-Resolution simulation results. The distributions agree between both simulations, except for the slow and low-density gas population, which is not refined due to its long cooling length.

\begin{figure}
    \centering
    \includegraphics[width=\linewidth]{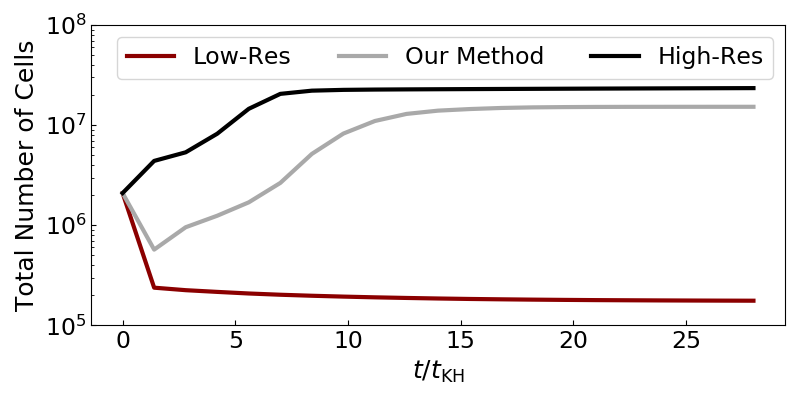}
    \caption{Total number of cells for the three Kelvin-Helmholtz test simulations. Our Method refines cells extensively after $t > t_{\rm KH} \approx 0.32$ Myr, rapidly increasing the total cell number, roughly matching the High-Resolution simulation value. By $15 \ t_{\rm KH}$, simulations performed with Our Method have about half the number of cells in the High-Resolution simulation.}
    \label{fig:refinement-test-KH-time}
\end{figure}

Finally, Figure~\ref{fig:refinement-test-KH-time} shows the time evolution of the number of gas cells in our three simulations.
The impact of our extra-refinement approach is evident, with the number of cells in Our Method (shown with the grey line) increasing by an order of magnitude from $1 t_{\rm KH}$ to $15 t_{\rm KH}$, matching the high-resolution simulation, depicted by the black line. The inflection point around $t_{\rm KH}$ marks where the central gas layer, initially cooler, heats up upon interacting with the surrounding hotter gas and begins to satisfy the refinement criteria.

Despite not perfectly matching the High-Resolution simulations (Figure \ref{fig:refinement-test-KH-histogram}), selectively maintaining higher resolution in the region of interest greatly enhances results over those obtained with Low-Resolution setups, showing similar behaviour as in High-Resolution (Figures \ref{fig:refinement-test-KH}, \ref{fig:refinement-test-KH-histogram}).
The effectiveness of our extra refinement method in capturing features of High-Resolution simulation at reduced computational costs is illustrated in Table \ref{tab:KH-cost-comparison}.
Simulations with additional refinement were run approximately $\approx 3$ times faster than their High-Resolution counterparts.
Though this improvement may depend on the simulation setup and focus, we actually find similar speed-up in our clumpy disc simulations (see Section ~\ref{subsec:new-sims})

\begin{table}[b]
\centering
\begin{tabular}{l|ccc}
\hline \hline
\textbf{Simulation}$^A$ & \textbf{L45\_M} & \textbf{L45\_R64} & \textbf{L45\_M\_HR} \\ \hline
\textbf{Target Mass ($\msun$)}$^B$ & 100 & 100 / 1.6$^*$ & 10 \\
\textbf{Cpuh to reach 1Myr}$^C$ & $5 \times 10^4$ & $3 \times 10^5$ & $1 \times 10^6$ \\
\textbf{Speed Ratio}$^D$ & 20x & 3.3x & 1x \\ \hline \hline
\end{tabular}
\caption{Comparison of computational cost across three of our simulations. \textbf{A:} Simulation name, same as in Table \ref{tab:simulations-params}. \textbf{B:} Simulation target mass. \textbf{C:} CPU hours spent to simulate up to 1Myr. \textbf{D:} Speed ratio, normalized to the high-resolution model. 
\newline \small{* The base target mass is $100 \ \msun$, for the cool outflowing gas, we reduced the target mass to $1.6 \ \msun$.}}
\label{tab:our-simulations-cost-comparison}
\end{table}

\subsection{Our simulation suite}\label{subsec:new-sims}

\begin{table*}[b]
\centering
\begin{tabular}{c|c|cccccccc}
\multicolumn{1}{c|}{} & \makecell{\textbf{Simulation}$^A$} & \makecell{$\lambda^B$ \\ (pc)} & \makecell{L$_{\rm AGN}^C$ \\ (erg s$^{-1}$)} & \makecell{$M_{\rm outflow \ cell}^D$ \\ ($\msun$)} & \makecell{$v_{\rm wind}^E$ \\ (km s$^{-1}$)} & \makecell{$\beta^F$} & \makecell{Duration$^G$ \\ (Myr)}  & \makecell{$T_{\rm floor}^H$ (K)} & \makecell{Subset$^I$} \\ 
\hline \hline

\multirow{10}{*}{\rotatebox[origin=c]{90}{Sims1 (From \textit{W24})}}
& L45\_L                & 333  & $10^{45}$   & 100 & $10^4$           & 1  & 6.1 & $10^4$ & Sims1-ISM \\
& L43\_M               & 167  & $10^{43}$   & 100 & $10^4$           & 1  & 15.7 & $10^4$ & Sims1-Lum \\
& L44\_M               & 167  & $10^{44}$   & 100 & $10^4$           & 1  & 5.3 & $10^4$  & Sims1-Lum \\
& L45\_M               & 167  & $10^{45}$   & 100 & $10^4$           & 1  & 5.2 & $10^4$  & Sims1-ISM / Sims1-Lum \\
& L45\_M\_HR           & 167  & $10^{45}$   & 10  & $10^4$           & 1  & 5.2 & $10^4$  & Sims1-ISM / Sims1-Lum \\
& L45\_M\_slowerWind   & 167  & $10^{45}$   & 100 & $5\times 10^3$   & 1  & 5.2 & $10^4$  & Sims1-ISM \\
& L45\_M\_fasterWind   & 167  & $10^{45}$   & 100 & $3\times 10^4$   & 1  & 5.2  & $10^4$ & Sims1-ISM \\
& L46\_M               & 167  & $10^{46}$   & 100 & $10^4$           & 1  & 1.4 & $10^4$  & Sims1-Lum \\
& L47\_M               & 167  & $10^{47}$   & 100 & $10^4$           & 1  & 1.0 & $10^4$  & Sims1-Lum \\
& L45\_S               & 40   & $10^{45}$   & 100 & $10^4$           & 1  & 5.0 & $10^4$  & Sims1-ISM \\
\hline \hline

\multirow{11}{*}{\rotatebox[origin=c]{90}{Sims2 (New Set)}}
& L43\_R16             & 167  & $10^{43}$   & 6.3 & $10^4$           & 16 & 1.0 & $10^4$  & Sims2-R16 \\
& L44\_R16             & 167  & $10^{44}$   & 6.3 & $10^4$           & 16 & 1.0 & $10^4$  & Sims2-R16 \\
& L45\_R16             & 167  & $10^{45}$   & 6.3 & $10^4$           & 16 & 1.0 & $10^4$  & Sims2-R16 \\
& L46\_R16             & 167  & $10^{46}$   & 6.3 & $10^4$           & 16 & 1.0 & $10^4$  & Sims2-R16 \\
& L47\_R16             & 167  & $10^{47}$   & 6.3 & $10^4$           & 16 & 1.0 & $10^4$  & Sims2-R16 \\
& L43\_R64             & 167  & $10^{43}$   & 1.6 & $10^4$           & 64 & 1.0 & $10^4$  & Sims2-R64 \\
& L44\_R64             & 167  & $10^{44}$   & 1.6 & $10^4$           & 64 & 1.0 & $10^4$  & Sims2-R64 \\
& L45\_R64             & 167  & $10^{45}$   & 1.6 & $10^4$           & 64 & 1.0 & $10^4$  & Sims2-R64 \\
& L46\_R64             & 167  & $10^{46}$   & 1.6 & $10^4$           & 64 & 1.0 & $10^4$  & Sims2-R64 \\
& L46.5\_R64           & 167  & $10^{46.5}$ & 1.6 & $10^4$           & 64 & 1.0  & $10^4$ & Sims2-R64 \\
& L47\_R64             & 167  & $10^{47}$   & 1.6 & $10^4$           & 64 & 1.0 & $10^4$  & Sims2-R64 \\

& L44\_ML\_T2             & 167  & $10^{44}$   & 100 & $10^4$           & 1 & 0.5 & $10^2$  & Sims2-Metal\\
& L44\_ML\_T4             & 167  & $10^{44}$   & 100 & $10^4$           & 1 & 0.5 & $10^4$  & Sims2-Metal \\
& L46\_ML\_T2             & 167  & $10^{46}$   & 100 & $10^4$           & 1 & 0.3 & $10^2$  & Sims2-Metal \\
& L46\_ML\_T4             & 167  & $10^{46}$   & 100 & $10^4$           & 1 & 0.3  & $10^4$ & Sims2-Metal \\

\hline \hline
\end{tabular}
\caption{Set of analysed simulations and their initial parameters. The columns, in order, represents: A) The simulation ID; B) Average largest cloud size in the initial ISM; C) The AGN luminosity; D) Target mass resolution for the extra refined gas (see section \ref{subsec:high-res-sims}). The base target mass is $100\beta^{-1} \msun$ for all simulations except L\_45\_HR; E) Injected wind velocity at the central boundary; F) Resolution boost factor (see equation \eqref{eq:refinement-criteria}); G) Duration of the simulation, in our analysis we usually looked at two selected times: 1 Myr and 4 Myr; H) The temperature floor; I) Simulation subset, see Section \ref{subsec:new-sims} --the group Sims2-Metal refer to simulations with metal line cooling present.
}
\label{tab:simulations-params}
\end{table*}

We use simulations from \textit{W24} and perform a series of new simulations with the refinement scheme outlined in Section~\ref{subsec:high-res-sims}.
Our new simulations, performed with additional refinement (``Our Method''), are all initialised with medium clumps (i.e., $\lambda = 167$pc; Section \ref{subsec:sims-data} and Table \ref{tab:simulations-params}).
This choice follows our finding \citep{Ward2024} that the size of ISM clumps introduces only second-order differences in the state of the fast-moving cool gas, a result we revisit in Section \ref{sec:results}.
In our new simulations, we employ two different refinement boost factors of $16$ or $64$ (see Section \ref{subsec:high-res-sims}), reducing the minimum target mass of the simulations to approximately $6.3 \, \rm M_\odot$ and $1.6 \, \rm M_\odot$, respectively. This resolution allows us to probe cool cloudlets beyond even the high-resolution achieved in \textit{W24}.

When comparing simulations with different resolutions, e.g. simulation L45\_M, L45\_B64, and simulation L45\_M\_HR (we discuss our simulation naming convention below), we find our refinement scheme provides a similar speed-up as in the Kelvin-Helmholtz test presented in Section~\ref{subsubsec:refinement-tests}.
Table \ref{tab:our-simulations-cost-comparison} lists the run-time and speed-up found in our new simulations. The computing time for the fiducial simulation (L45\_M) and the high-resolution simulation (L45\_M\_HR) from \textit{W24}, which differ in their target mass by a factor of $10$, differs by a factor $20$.
Our highest-resolution simulation (L45\_B64), with a resolution boost of $64$ in the outflowing gas, consumes $3.3$ times less computational time than L45\_M\_HR, despite a smaller target mass for outflowing, cooling gas.

A list of all the simulations used in this paper and their main parameters is provided in Table \ref{tab:simulations-params}.
The simulations are categorised into two groups: those from \textit{W24} (denoted as Sims1) and our new simulations featuring extra refinement (denoted as Sims2). Within Sims1, we distinguish two subgroups: Sims1-Lum, which explore different values for $L_{\rm AGN}$ at fixed initial ISM condition (medium clump sizes), and Sims1-ISM, which vary initial ISM conditions and injected wind velocity at a fixed $L_{\rm AGN} = 10^{45}$ erg s$^{-1}$.
Simulations Sims2 probe two resolution boost factors for fast-moving cool gas across various $\lagn$ values (Sims2-R16 and Sims2-R64). We explore the effects of metal-line cooling in Sims2-Metal. Most of our analysis will work using the set with highest resolution, Sims2-R64, exploring the impact of varying $\lagn$ at a fixed boost factor.
Our simulation naming convention indicates the AGN luminosity (e.g., L45 refers to $\lagn = 10^{45}$ erg s$^{-1}$), followed by a letter that corresponds to the initial ISM model (for simulations from \textit{W24}), to the refinement boost factor (for the new simulations),or ML\_T for metal-line cooling simulations (the temperature floor following the T: $2$ for $10^2$~K or $4$ for $10^4$~K). For the \textit{W24} suite, `S', `M' and `L' denote simulations with small, medium and large average cloud sizes, respectively (see second column of Table \ref{tab:simulations-params} for the corresponding cloud size). Additionally, three simulations include an additional suffix indicating a special feature: L45\_M\_HR is the original high-resolution simulation from \textit{W24}, with $M_{\rm target} \, = \, 10 \, \rm M_\odot$; L45\_M\_slowerWind and L45\_M\_fasterWind are simulations with different AGN wind velocity of $v_{\rm wind} \, = \, 5 \times 10^3$ km s$^{-1}$ and $v_{\rm wind} \, = \, 3 \times 10^4$ km s$^{-1}$ respectively. The total duration of each simulation varies, ranging from 1 to 16 Myr, as shown in Table~\ref{tab:simulations-params}; for more details, refer to \textit{W24}.

\subsection{Identifying cool clouds in the outflow}\label{subsec:find-clouds}

\textit{W24} showed that cooling in turbulent mixing layers between initial cool clumps and AGN wind fluid proceeds on a short time-scale, giving rise to a population of cool, outflowing clouds.

In this study, we investigate the properties of these cool cloudlets and their relation to the AGN wind properties.
To extract cool clouds from our simulations, we employ the DBSCAN clustering algorithm \citep{Ester1996, Khan2014, Schubert2017, Deng2020}, known for its capability to detect clusters of arbitrary shapes and sizes. DBSCAN categorizes data points based on proximity, defining a cluster as a collection of points connected within a specified distance $r_{\rm DB}$ and meeting predefined criteria (e.g. minimum number of elements, minimum density of points). This method has been used on astronomical data \citep{Tramacere2013, Zhang2019, Logan2020, Prisinzano2022, Raja2024}, being especially useful to identify cluster elements and categorization.

We only consider cool gas cells, i.e. with temperature below $2 \times 10^4$ K. Two cells are considered neighbours if their positions are less than $r_{\rm DB}=4\overline{D}$ apart, where $\overline{D}$ is the average size of cells with temperature lower than $2 \times 10^4$ K (see Appendix \ref{app:DBSCAN-tests}). Additionally, to ensure statistical significance, we only consider clouds if they contain at least $100$ member cells.
To isolate \emph{outflowing} clouds, we consider only CGCs with average radial velocities exceeding 10 km s$^{-1}$ -- the same velocity cut used in \textit{W24} (see Section 2.6 in \textit{W24}). Since our ISM model represents an idealized scenario that excludes rotation or velocity dispersion, any gas motion is attributable to AGN activity. Radiative cooling introduces small pressure gradients that induce some gas motion, but, as shown by \textit{W24}, applying a velocity threshold of $10$ km s$^{-1}$ filters these out very effectively.

\begin{figure}
    \centering
    \includegraphics[width=\linewidth]{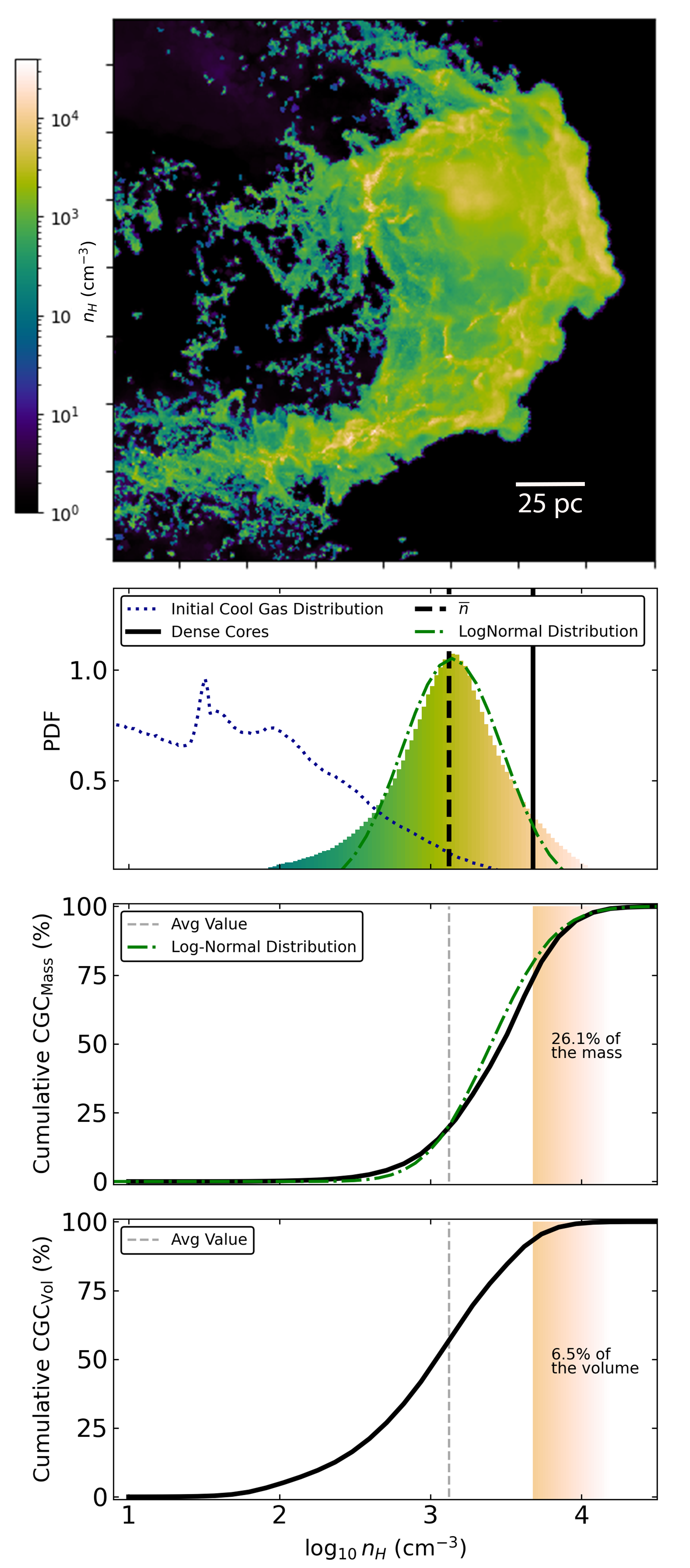}
    \caption{The top panel shows a density-weighted projection with a depth of 100 pc, centred on a massive outflowing cool gas cloud (CGC) with $M_{\rm CGC} \sim 10^{5} \ \msun$, extracted from simulation L45\_R64. This density map highlights its complex internal structure, with an internal density ranging from $1$ cm$^{-3}$ to $3 \times 10^{4}$ cm$^{-3}$. For the same CGC, we show the density distribution (second row), cumulative mass distribution (third row), and cumulative volume distribution (last row). The density distribution follows an approximately log-normal distribution (green dash-dotted curve). The dotted blue curve shows the initial cool gas distribution, for comparison. In the third and fourth panels, the coloured area highlights the CGC cores, comprising the top 10\% densest cells that account for more than 25\% of the total CGC mass comprised in less than 7\% of the total CGC volume.}
    \label{fig:CGC-Density-Distr}
\end{figure}

The physical properties of each CGC, such as temperature or velocity, are determined using the mass-weighted average of all its member gas cells. For instance, a cloud's velocity is calculated as
\begin{equation}
    \vec{v}_{\rm cloud} = \frac{\sum m_{\rm (CGC \ cells)} \times \vec{v}_{\rm (CGC \ cells)}}{\sum m_{\rm (CGC \ cells)}} \, .
    \label{eq:mass-weighted-average}
\end{equation}
Other physical quantities are estimated similarly.
For simplicity, we calculate the size of a CGC, denoted as $R_{\rm cloud}$, as the radius of a sphere of identical volume, though we note that CGCs are not spherical (e.g. see Figure~\ref{fig:CGC-Density-Distr}).

Within each CGC, we find a complex density structure, with density variations spanning up to four orders of magnitude. While most cells within these clouds exhibit densities close to the cool phase average of $\approx 5$ cm$^{-3}$, there are regions where the density peaks at values exceeding $10^4$ cm$^{-3}$, as illustrated in Figure \ref{fig:CGC-Density-Distr}, which presents a density-weighted density projection ($100$ pc depth) of a large CGC with $M_{\rm CGC} \sim 10^5 \ \msun$, extracted from L45\_R64. Most gas has a density between $10^{2.5}$ and $10^{3.5}$ cm$^{-3}$, while smaller regions contain extremely high-density peaks exceeding $10^{3.5}$ cm$^{-3}$.

In Section~\ref{subsec:obs-comparison}, we investigate whether these high-density regions behave differently from the average CGC gas, hypothesizing that they are responsible for the observed emissions of several emission lines (e.g. [S~{\sc ii}], [O~{\sc iii}]).
We thus define the {\em cores} of CGCs as the regions containing the top 10\% densest cells. The density of these cores varies from cloud to cloud and, as we go on to show, scales with $\lagn$.
The dense cores can account for 10--30\% of the total mass of a CGC. A single CGC, especially the largest and most massive ones, can contain multiple cores, as illustrated in the first panel of Figure \ref{fig:CGC-Density-Distr}. Subsequent panels of Figure \ref{fig:CGC-Density-Distr} present the density distribution, cumulative mass and volume distribution of gas within a typical CGC.
This distribution is shown with the green curves in the second and third panels of Figure \ref{fig:CGC-Density-Distr}, which can be compared by the actual density distribution (the shaded histograms in the second panel and the black curve in the two bottom panels). We note that the final CGC distribution is very different from the initial cool gas density distribution, which is shown by the dotted blue line in the second panel.
Small clouds with $n_{\rm H, cells} \sim 100$ exhibit noisy density PDFs, whereas more massive CGCs ($M_{\rm CGC} > 10^3$~M$_\odot$) show an approximately lognormal shape with enhanced wings, as illustrated for a representative cloud in the second panel of Figure~\ref{fig:CGC-Density-Distr}. The best-fitting lognormal parameters vary with several factors (distance, total CGC mass, and $L_{\rm AGN}$). Many cloud-crushing studies (e.g. \citealt{Federrath2010, Federrath2013, Mukherjee2016, Mukherjee2018}) report density PDFs that are well described by an approximate lognormal with a time-dependent power-law tail. The internal density distributions we find are close to lognormal, in tantalising agreement with these works. A more detailed investigation of these properties is promising, but would require a more realistic setup including self-gravity, magnetic fields, and other physics that can affect the internal cloud structure.

The bottom two panels highlight the contrast between the core densities and the overall average density of the CGC, demonstrating the significant mass contribution from these dense regions, which occupy only a small fraction of the CGC's total volume.

We have conducted extensive tests to our cloud extraction method. Each inspection confirmed the presence of gas with temperatures below $2 \times 10 ^4$ K and number densities exceeding $10$ cm$^{-3}$ for the member cells.  Through the IDs of randomly selected CGC cells, we have also verified that detected clouds consist of \texttt{AREPO} cells that can be linked through neighbour chains.
It was observed that the CGCs identified do not encompass all the cool gas within the simulations. Due to our selection criteria, and in particular the requirement for a minimum cluster size of $100$ cells, some cool gas cells were excluded from any CGC.
However, in Section \ref{subsec:AGN-Gas-Impact} we also show that many of our results are, in fact, independent of how we group cells into clouds. A more detailed set of tests involving DBSCAN and the two free parameters used by the model is presented in Appendix \ref{app:DBSCAN-tests}.

We also verified if our results are robust to the choice of temperature threshold used to define the cool phase. We analysed the CGCs using a slightly higher temperature cut of $T < 10^5$K, and we found no significant difference in the results. Due to the efficient cooling at this temperature in our simulations, there is very little gas with temperatures between $2 \times 10^4~$ and $5 \times 10^5$~K, and the bulk of the cool material lies below either threshold. This confirms that our findings on the properties of outflowing clouds are not sensitive to whether the cut is placed at $2 \times 10^4$K or $10^5$K.

\section{Results}\label{sec:results}

\begin{figure*}
    \centering
    \includegraphics[width=.95\linewidth]{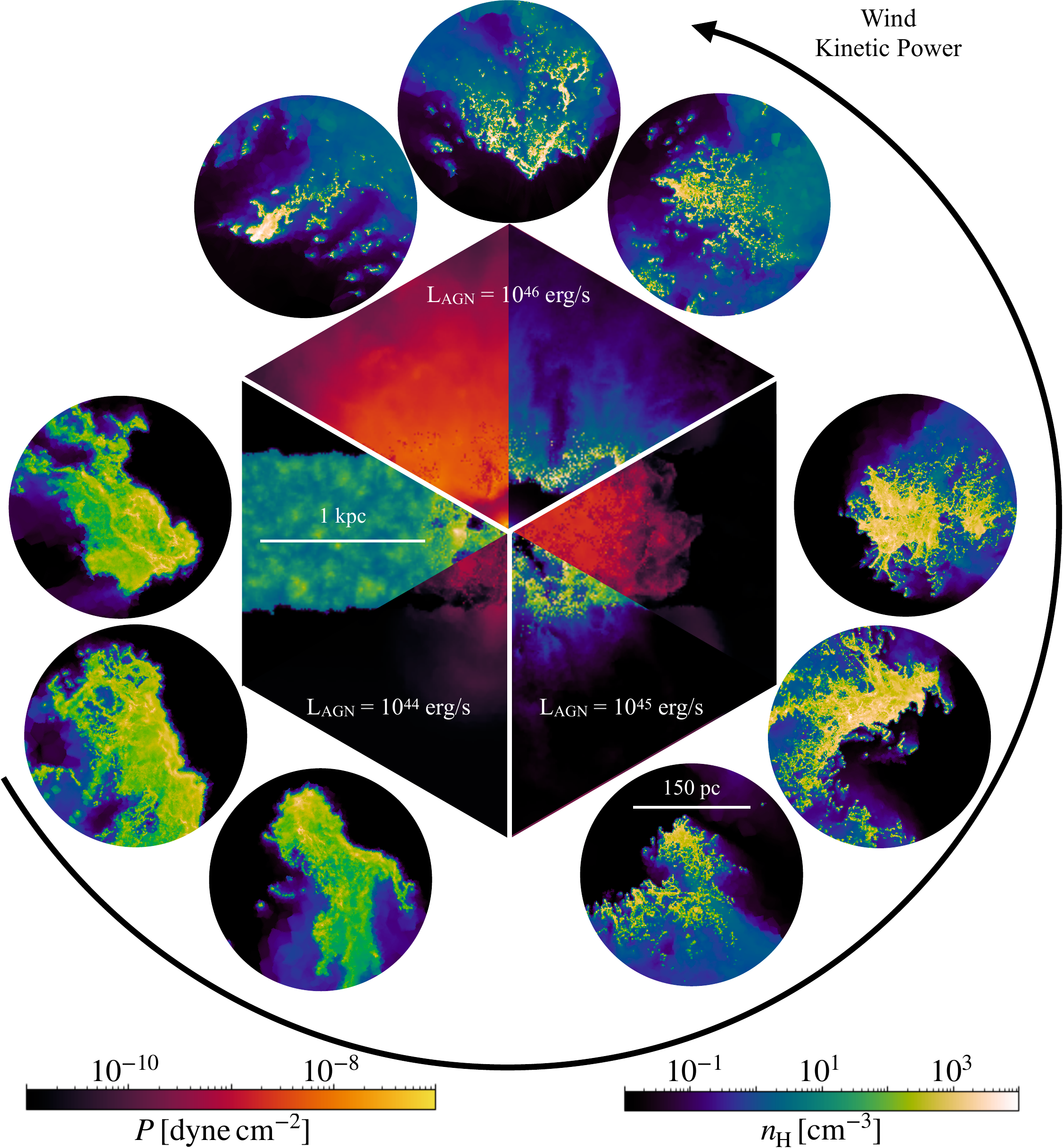}
    \caption[]{Density-weighted projections showing gas pressure (red/orange colour) and density (blue/green colour) for simulations with $\lagn \, = \, 10^{44} \, \rm erg \, s^{-1}$ (bottom, left section), $\lagn \, = \, 10^{45} \, \rm erg \, s^{-1}$ (bottom, right section) and $\lagn \, = \, 10^{46} \, \rm erg \, s^{-1}$ (top section). The circular inset plots surrounding the central panel show density projections that zoom-in on some of the outflowing clouds found in these simulations. As the AGN luminosity and wind kinetic power increase, outflowing clouds become denser as their surrounding pressure grows, and shatter into smaller cloudlets. The spatial scale is the same for all inset plots.}
    \label{fig:maps}
\end{figure*}

We present our results in this Section.
We start in Section \ref{subsec:Global-Impact} with qualitative overview of the main results of this paper. In Section \ref{subsec:Impact-Initial-ISM}, we assess the influence of initial ISM conditions on the outflowing cloud population. We investigate the effects of $\lagn$ on the demographics and internal structure of cool outflowing clouds in Sections \ref{subsec:Impact-LAGN}, and their time evolution in \ref{subsec:temporal-evolution}. In Section \ref{subsec:AGN-Gas-Impact}, we generalize our findings, demonstrating that our main results hold irrespective of how CGCs are defined.

\subsection{From large ISM clumps to fine, outflowing cloudlets}\label{subsec:Global-Impact}

Figure \ref{fig:maps} presents an overview of our simulations. The central hexagon is separated into three different sections, corresponding to simulations L44\_R64 (bottom, left section), L45\_R64 (bottom, right section) and L46\_R64 (top section). Each section shows projected, density-weighted gas pressure and density for the central few kiloparsec of the simulated clumpy disc.

\begin{figure*}
    \centering
    \includegraphics[width=\linewidth]{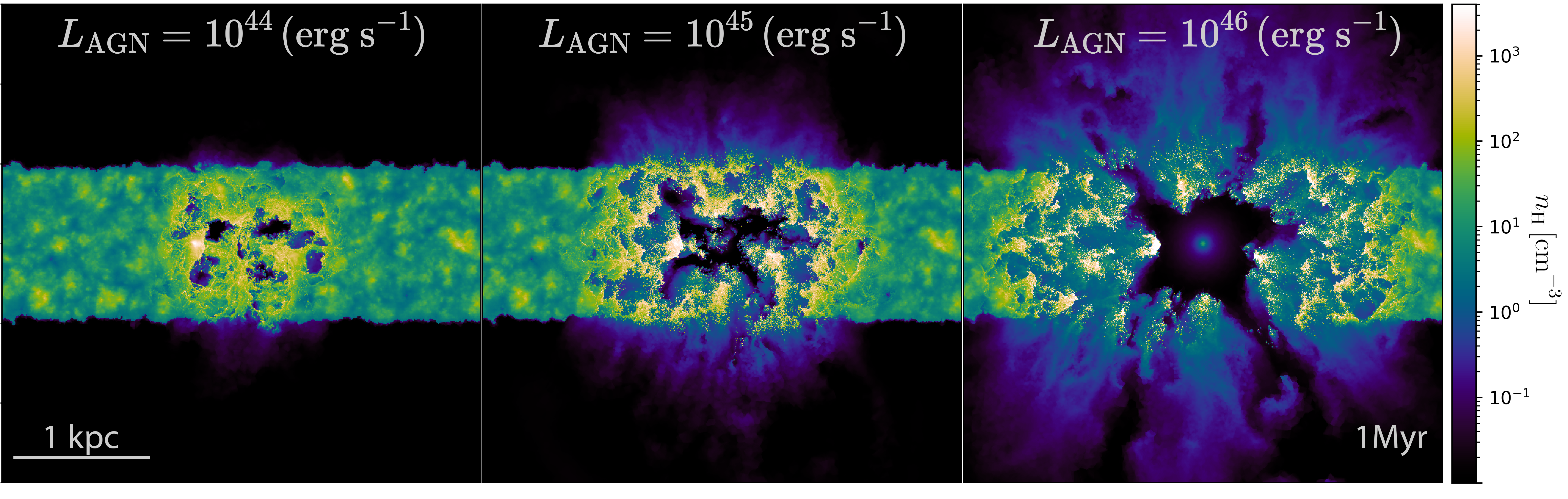}
    \caption[]{Density-weighted projections of gas density for simulations with $\lagn = 10^{44} \, \rm erg \, s^{-1}$ (left), $\lagn = 10^{45} \, \rm erg \, s^{-1}$ (centre), and $\lagn = 10^{46} \, \rm erg \, s^{-1}$ (right) at $t = 1$~Myr. At a given time, brighter AGN power more energetic large-scale outflows. These clear out gas from the galactic nucleus, and eventually break out of the galactic disc, creating a kpc-scale bubble that can extend beyond the disc.}
    \label{fig:rho-maps}
\end{figure*}

The impact of the wind is particularly clear for L45\_64 and L46\_64, as shown in Figure \ref{fig:rho-maps}. This illustrates a kpc-scale density-weighted projection of gas number density, highlighting the large-scale interaction between the AGN wind and the surrounding ambient medium. More powerful AGN produce larger low-density cavities in the galactic centre, inflating prominent kpc-scale bubbles \citep{Ward2024}. In L46\_64, the more energetic AGN wind carves out a particularly clear cavity. At larger scales, ISM clumps remain mostly intact in L44\_M, but are strongly disrupted in L45\_M, and L46\_M, where they can be seen to fragment into finer cloudlets. Behind these cloudlets, aligned with the wind, are long wakes containing ram pressure-stripped clump material. These wakes extend out to a few kiloparsec into the halo in L46\_M, raising the ambient density outside the disc.

The maximum pressure and the size of the high-pressure bubble can also be seen to increase with AGN luminosity in Figure \ref{fig:maps}. This is because the mass flux of the small-scale wind scales linearly with $L_{\rm AGN}$, resulting in a post-shock pressure that scales as $L_{\rm AGN}$ (Section~\ref{subsec:physical-interpretation}). The pressure confining the fine outflowing clouds thus rises with AGN luminosity. Careful inspection of the pressure maps further reveals these cloudlets are mildly under-pressurised with respect to the diffuse outflow component, which is often seen in simulations of wind-cloud interaction \citep[e.g.][]{Fielding2020}.

The circular panels surrounding the central hexagon in Figure \ref{fig:maps} ``zoom-in'' on a few of these fine, outflowing cloudlets, revealing their internal structure in striking detail.
As already shown in Section~\ref{subsec:find-clouds}, these cloudlets are highly inhomogeneous, with internal densities spanning several orders of magnitude.
For fixed $L_{\rm AGN}$, these clouds have similar density distributions, sizes and morphologies.
\emph{However, we find that all these properties change systematically with AGN luminosity}. With higher AGN luminosity, we see that (i) clouds break down into smaller fragments, and (ii) the overall density of these fine clouds increases. While all cloudlets contain high density cores (see also Section~\ref{subsec:find-clouds}), their density also appears to scale with AGN luminosity.
Note also that the overall cloud densities are $2$ to $3$ orders of magnitude greater than the densities in the initial clumps.

In our simulations, we set the small-scale wind properties at injection, but have no control over its interaction with the ambient gas. The large-scale outflow and the properties of fine, outflowing cloudlets and their scaling with AGN luminosity thus constitute a genuine, theoretical prediction of our simulations. In the following sections, we investigate more quantitatively how outflowing cloudlet demographics, size and density vary with time and with AGN luminosity.

\subsection{Impact of initial clump size and small-scale wind velocity}\label{subsec:Impact-Initial-ISM}

\begin{figure*}
    \centering
    \includegraphics[width=\linewidth,trim={.0cm 0cm 0.cm 0cm},clip]{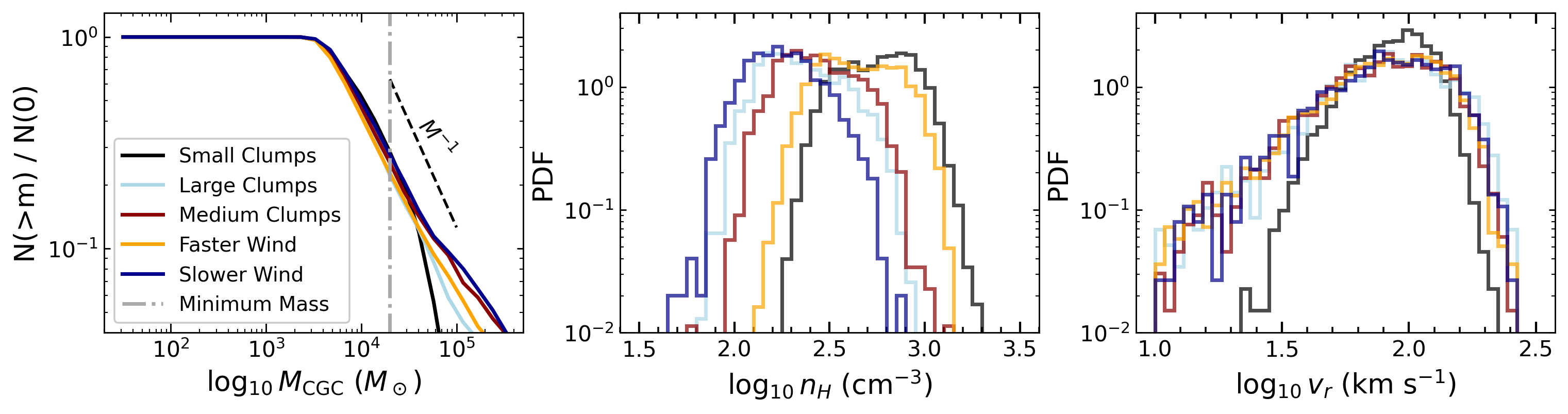}
    \caption{Comparison of simulations with different initial ISM clump sizes --small, medium, and large-- and different AGN wind velocities --$v_{\rm wind}=5 \times 10^3$ km s$^{-1}$ (Slower), $v_{\rm wind}=10^4$ km s$^{-1}$, and $v_{\rm wind}=3 \times 10^4$ km s$^{-1}$ (Faster). \textit{Left:} Mass distribution for clouds above a certain mass threshold, showing a slope of $\sim -1$ in our simulations, indicative of a mass distribution $dN/dM \propto M^{-2}$. The grey line indicates where clouds enclose less than 100 cells. \textit{Center:} Density distribution for the same CGCs, highlighting systematic deviations where smaller clumps and faster winds produce denser CGCs, by a factor of less than $2$. \textit{Right:} Velocity distribution for the same CGCs. On the higher velocity side $v_r > 100$ km s$^{-1}$, all simulations look the same. For $v_r < 100$ km s$^{-1}$, simulations with large and medium clumps yield similar velocity distributions, whereas smaller clumps facilitate higher minimum velocity CGCs.}
    \label{fig:k-hist-t41}
\end{figure*}

The left-hand panel of Figure \ref{fig:k-hist-t41} presents the cumulative mass distribution of outflowing CGCs at 4 Myr, normalised to the total mass in CGCs.
The distributions are shown for Sims1-ISM suite of simulations (see Section~\ref{subsec:new-sims}), i.e. for the three different initial ISM cloud distributions and a fixed $L_{\rm AGN} \, = \, 10^{45} \, \rm erg \, s^{-1}$.
This panel includes all identified CGCs, regardless of number of constituent cells. For subsequent analysis, only CGCs above a minimum threshold of $100$ constituent cells (grey, dot-dashed line) are considered (see Section \ref{subsec:find-clouds}), ensuring each cloud is sampled by sufficient resolution elements.

By $4$ Myr, the cumulative CGC mass distributions settle into a power law in all simulations, regardless of initial ISM clump size. The power law index is $\approx -1$ for well-resolved CGCs, with masses larger than $\sim 10^{3} \ \msun$, and varies only mildly between $\approx -0.7$ (L45\_S) to $\approx -1.1$ (L45\_M\_slowerWind).
As we show in Sections \ref{subsec:Impact-LAGN} and \ref{subsec:temporal-evolution}, the power law index is robust against AGN luminosity variations and time evolution.
In fact, the slope $\alpha \approx -1$ in the cumulative mass distribution $N$ gives $dN/dM \propto M^{-1+\alpha} = M^{-2}$.
The mass distribution of surviving, outflowing clouds in our simulations is in fact consistent with a radiative, turbulent mixing layer origin \citep{Gronke2022, Tan2024, Colman2024, Warren2024}.

The central panel of Figure \ref{fig:k-hist-t41} presents distributions of the mean internal number density of CGCs at 4 Myr for different simulations, where the mean values are computed for individual CGCs using a mass-weighted average as in Eq.~\ref{eq:mass-weighted-average}. This panel thus gives the distribution of mean internal density of CGCs -- not the number density of CGCs in the simulation. We see quantitative differences between simulations with different initial ISM conditions, with the peak of the density distribution varying from $\approx 200 \, \rm cm^{-3}$ to $\approx 1000 \, \rm cm^{-3}$.

At fixed small-scale wind velocity, simulations with initially larger clumps produce CGCs with the lowest average density. The most distinct behaviour is observed in the Small Clumps simulation (black line), where the densest clouds are produced, with average densities $\approx 0.3$ dex higher than the other simulations with the same wind velocity.

\begin{figure*}
    \centering
    \includegraphics[width=\linewidth]{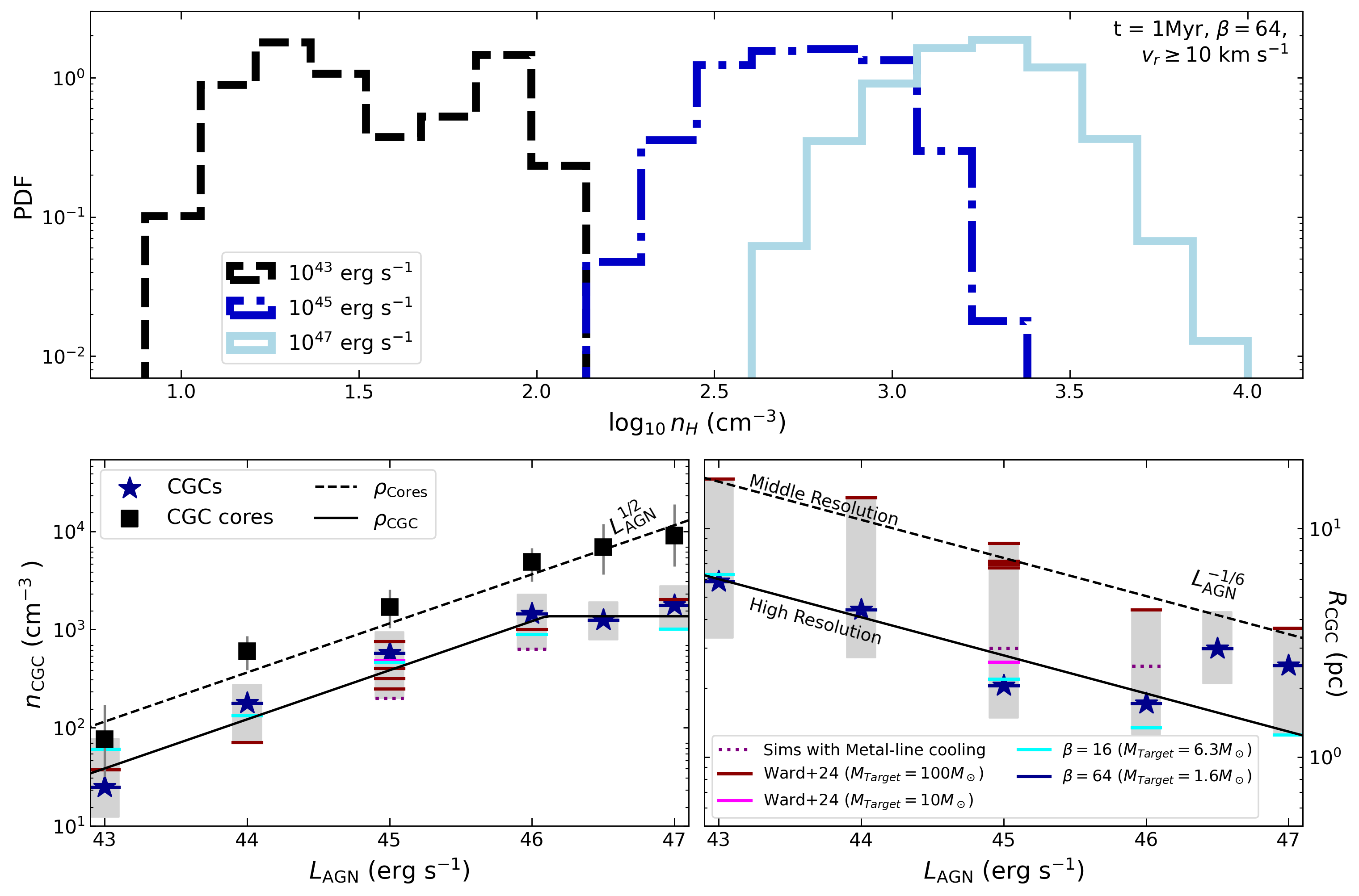}
    \caption[]{Examining the impact of AGN luminosity on the main characteristics of the outflowing CGCs. (\textit{Top panel}) The density distributions from our highest resolution simulations for $\lagn  = 10^{43}, \ 10^{45}, \text{ and } 10^{47}$ erg s$^{-1}$ at 1 Myr, showing a shift to higher values with increasing AGN power. The lower panels illustrate the average density of CGCs (left) and the average size of the clouds (right), as a function of the AGN bolometric luminosity for a range of simulations investigated in this work at $t = 1$~Myr (except for the simulations with metal-line cooling, which the time is the maximum simulation time shown in Table \ref{tab:simulations-params}). The main results, showed as the dark blue stars and black squares, refer to the highest resolution simulations group --with resolution boost factor $\beta = 64$.
    The shaded grey areas highlight the full range of values across all of the simulations (see Table \ref{tab:simulations-params}), plus an 1$\sigma$ error range from the main run. The average values from the different simulations are shown as individual horizontal lines as described in the legend. In the lower left panel, average density values for the CGC cores from the main simulations are also displayed. (\textit{Lower Left Panel}): We can see the CGC's density increasing with $L_{\rm AGN}^{1/2}$, up to the breakpoint for $\lagn \geq 10^{46}$ erg s$^{-1}$. This trend is consistent across all our simulations. However, this breakpoint is not observed when only the high density cores are analyzed (see secion \ref{subsec:Impact-LAGN}). (\textit{Lower Right Panel}): The average CGC sizes follow $L_{\rm AGN}^{-1/6}$.
    Our simulations set can be divided into Middle Resolution ($M_{\rm target} = 100 \ \msun$) and High Resolution ($M_{\rm target} = \lesssim 10 \ \msun$). The change in resolution did not affect the density results but did show a reduction in the average cloud size, up to the break at $\lagn = 10^{46}$ erg s$^{-1}$. Our results indicate that brighter AGNs result in smaller denser cool outflowing clouds.}
    \label{fig:L-hist}
\end{figure*}

A drop (increase) in wind speed corresponds to a proportional decrease (increase) in the small-scale wind power and in the mean hot gas pressure of the shocked wind bubble.
Looking at the two simulations where we vary wind velocity (at injection), the simulation with a slower wind, L45\_M\_slowerWind produces lower-density CGCs, with an mean density lower by about $0.2$ dex compared to the L45\_M simulation (dark red line). On the other hand, the simulation with a faster wind, L45\_M\_fasterWind produced denser CGCs, higher by $0.1$ dex when compared to L45\_M.

The right-hand panel of Figure \ref{fig:k-hist-t41} shows the radial velocity distribution of the CGCs, which spans a broad range from $10$ km s$^{-1}$ up to approximately $300$ km s$^{-1}$. While more diffuse cool gas can reach velocities $\approx 500 \, \rm km \, s^{-1}$ \citep{Ward2024}, this velocity is still much lower than the small-scale wind speed of $10^4$ km s$^{-1}$, a point we return to in Section \ref{subsubsec:velocity-interpretation}.
The velocity distributions for the simulations with Large and Medium Clumps are similar, and differences are only perceptible in the simulation with Small Clumps for $v_r \lesssim 100$ km s$^{-1}$, which shows a much steeper distribution.
In the Small Clumps simulation, the hot shocked wind bubble is confined more effectively by the cool ISM, mitigating escape along paths of least resistance, reducing the width of the velocity distribution \citep{Ward2024}.

Our findings suggest that the initial clump size has mild to moderate impact on the properties of CGCs, modulating their density by factors $\approx 2$, altering the low-velocity end of the distribution (Figure \ref{fig:k-hist-t41}).
As we show in the next section, these variations are small when compared to the effect of varying AGN luminosity.
For this reason, in the remainder of this paper, unless specified otherwise, we consider Medium Clump simulations and a small-scale wind velocity of $v_{\rm wind} = 10^4$ km s$^{-1}$.

\subsection{Impact of AGN luminosity and wind kinetic power}\label{subsec:Impact-LAGN}

We here focus on our highest-resolution simulations from the Sims2 suite (see Table \ref{tab:simulations-params}). These simulations start from medium-sized initial clumps, adopt a wind velocity of $v_{\rm wind} = 10^4$ km s$^{-1}$, and a resolution boost factor of $\beta = 64$ (check Section \ref{subsec:high-res-sims}).

The top panel of Figure \ref{fig:L-hist} shows the distribution of CGC internal density at $1 \, \rm Myr$ for $\lagn = 10^{43}, 10^{45}, \text{ and } 10^{47}$ erg s$^{-1}$. Simulations with $\lagn = 10^{44}$ and $10^{46}$ erg s$^{-1}$ are omitted for clarity, but they fit the same trend. As had been shown in Figure \ref{fig:maps}, it is evident that the CGC density distribution shifts to higher values with increasing AGN luminosity.
For instance, the density distribution for $\lagn = 10^{45}$ erg s$^{-1}$ is $\sim 10$ times higher than for $\lagn = 10^{43}$ erg s$^{-1}$, pointing to more efficient compression at higher AGN luminosities (see Section \ref{subsec:physical-interpretation}).

\begin{figure}
    \centering
    \includegraphics[width=\linewidth,trim={0cm 0cm 0cm 0cm},clip]{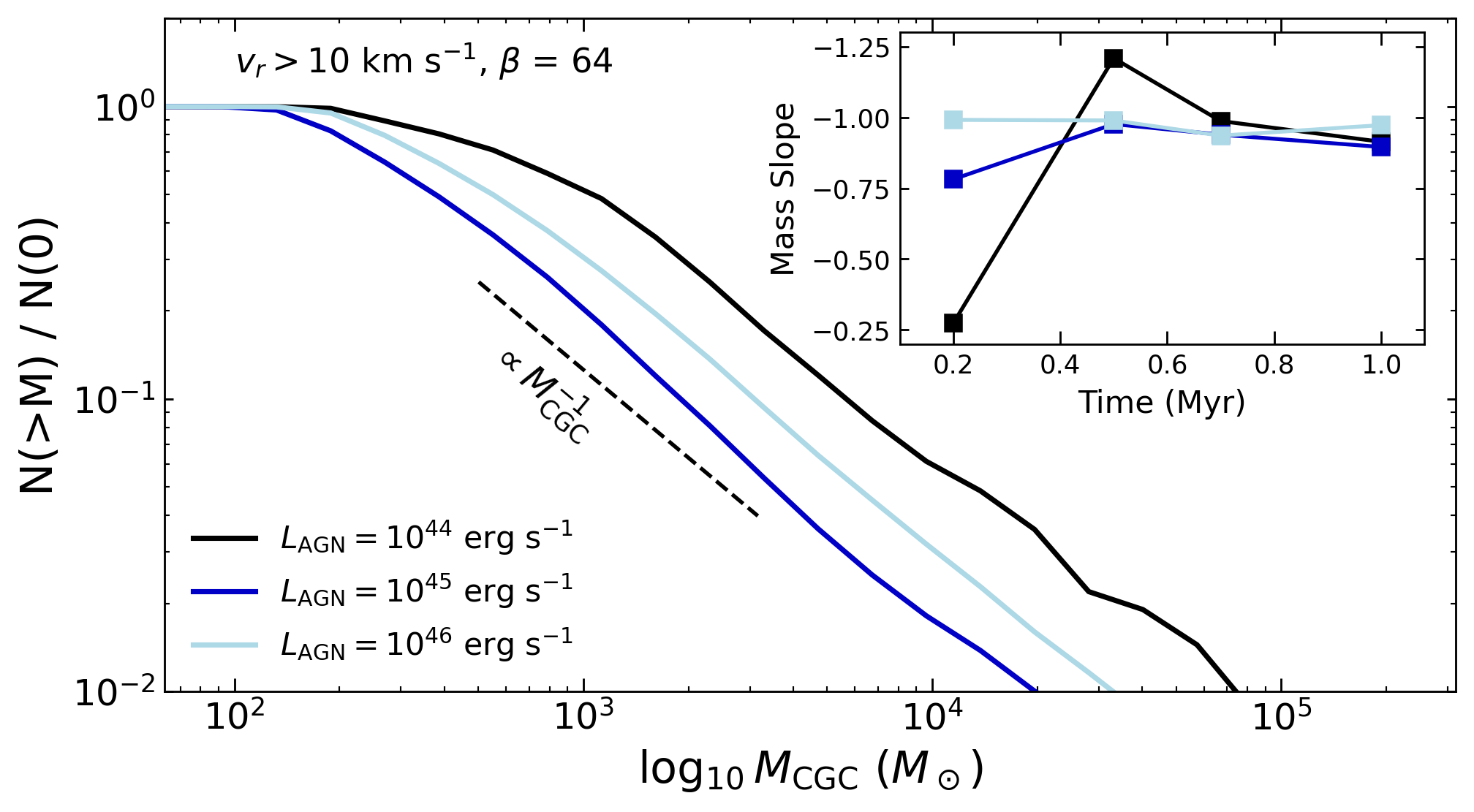}
    \caption[]{\textit{Main plot:} CGC cumulative mass distribution across different AGN power levels at $\mathbf{t = 1} \, \text{Myr}$, maintaining a consistent slope of $N(>M) \propto M^{-1}$, aligning with observations from Figure \ref{fig:k-hist-t41}. \textit{Inset:} The evolution of the slope over time, starting from $0$ at the inception of the simulations and adjusting to a range between $-0.7$ to $-1.2$, showing that the rate of change correlates directly with the AGN luminosity.}
    \label{fig:M-slope-time}
\end{figure}

The lower left-hand panel shows the mean CGC internal density in the highest-resolution simulations from the Sims2 suite as a function of AGN luminosity (dark blue stars). This panel shows a clear $\rho_{\rm CGC} \propto \lagn^{1/2}$ scaling.
The lower right-hand panel of Figure \ref{fig:L-hist} reveals the relation between $\lagn$ and the approximate physical size of CGCs, defined as $R_{\rm CGC} =  \frac{3}{4\pi}V_{\rm CGC}^{1/3}$, where $V_{\rm CGC}$ is the cloud volume. The sizes scale, approximately, as $R_{\rm CGC} \propto \lagn^{-1/6}$, though with much more significant scatter than seen in the density relation.

Both the density and size relations show a break at $\lagn \gtrsim 10^{46}$ erg s$^{-1}$. At these high luminosities, the average density stabilizes at approximately $10^3$ cm$^{-3}$.
This effect is caused by the more efficient cloud ablation with increased kinetic power (see Section \ref{subsec:physical-interpretation}). If we consider only the CGC cores (black squares in the lower left panel of Figure \ref{fig:L-hist}), no such break is seen and the same density and size relations hold out to the highest AGN luminosities, consistently following the $\lagn^{1/2}$ scaling, although with a $\approx 5$ times normalisation. Further discussion on this finding can be found in Section \ref{subsec:physical-interpretation}.

Increasing resolution changes neither the normalisation nor the slope of the CGCs' density scaling, as it is shown in the lower left-hand panel of Figure \ref{fig:L-hist}. Results for Sims1, the simulations from \textit{W24} with lower mass resolution, are shown with dark red ($M_{\rm target} = 100 \ \msun$) and magenta ($M_{\rm target} = 10 \ \msun$) lines, while Sims2, the new simulations with enhanced resolution, are shown dark blue ($M_{\rm target} = 100/64  \ \msun = 1.6 \ \msun$) and cyan ($M_{\rm target} = 100/16 \ \msun = 6.3 \ \msun$). Plotting the results for both groups, show that changing the resolution by a factor of up to $64$ does not change the mean CGC density values, suggesting that the density converges to a relation

\begin{equation}
n_{\rm CGC} \approx 50 \left(\frac{\lagn}{10^{43} \text{ erg s}^{-1}}\right)^{1/2} \mathrm{ cm}^{-3}.
    \label{eq:lagn_x_n}
\end{equation}
-- our $n_{\rm CGC}$ always refer to the hydrogen density.

Varying the resolution, however, does have a substantial impact on CGC sizes, as shown in the lower right-hand panel of Figure \ref{fig:L-hist}. Both the intermediate resolution simulations from Sims1, with a target mass for the outflowing cells of $M_{\rm OC} = 100 \, \rm \msun$, and the high-resolution simulations from Sims2 and L45\_M\_HR (see Table \ref{tab:simulations-params}), with a target mass of $M_{\rm OC} \lesssim 10 \ \msun$, display a similar trend, but with a different normalisation.
On average, the high resolution simulations produce CGCs that are about $2.5$ times smaller than their intermediate-resolution counterparts.
Though we might expect CGCs to shrink further in size at even higher resolution $\beta = 64$ (target mass of $1.6 \, \rm \msun$), our results suggest convergence in CGC properties for masses greater than $\approx 100 \msun$. We find the typical size-scale is $R_{\rm CGC} \approx 6 (\lagn / 10^{43} \text{ erg s}^{-1})^{-1/6}$ pc for the Sims2 simulations.
Examining the Sims1 simulations, we do not observe a break at $\lagn = 10^{46}$ erg s$^{-1}$, as seen in the CGC density plot. The average CGC size follows a relation $\propto \lagn^{-1/6}$. However, the Sims2 simulations, which achieve much higher resolution for gas with $T < 2 \times 10^4$K, do capture a break at the same luminosity as seen in the density relation. Both the $\beta = 16$ and $\beta = 64$ simulations, as well as the high-resolution run from Sims1, exhibit this break and show consistent normalisation in the size-luminosity relation. While the standard Sims1 simulations reproduce the correct average density, they lack the resolution required to resolve smaller cloudlets and identify the same deviation from the power-law at $\lagn = 10^{46}$~erg~s$^{-1}$. This analysis suggest extreme resolution is necessary to accurately measure the physical scales of these cloudlets, which are very small ($\sim 1$pc scales) and as we increase resolution, smaller ones can be detected.

\subsection{Time evolution of cool, outflowing cloud properties}\label{subsec:temporal-evolution}

\subsubsection{Cloud Mass Distribution}
In Figure \ref{fig:M-slope-time} we show the CGC mass distribution for the simulations with $\beta = 64$. Different colours represent different AGN luminosities. The observed power law for the CGCs in the mass range $10^2$ to $10^5 \msun$ at $1$ Myr has an index $\approx -1$ for all simulations shown, the same value found in Section \ref{subsec:Impact-Initial-ISM} for different initial ISM conditions.
The inset in Figure \ref{fig:M-slope-time} shows the time-evolution of the power law index for simulations with AGN luminosity $10^{44}$ (black), $10^{45}$ (blue), and $10^{46}$ erg s$^{-1}$ (light blue).
At the beginning, all gas has null $v_r$, and there are no CGCs. The first CGCs are formed at $0.1-0.2$ Myr, depending on the wind power.
At $1$ Myr, and at subsequent times, the power law index in all simulations converges to $\alpha \approx -1$. We observe that more powerful AGN winds set this relation faster due to their shorter outflow timescale, as also shown in the central panel of Figure \ref{fig:maps}.

\begin{figure}
    \centering
    \includegraphics[width=\linewidth,trim={0cm 0cm 0cm 0cm},clip]{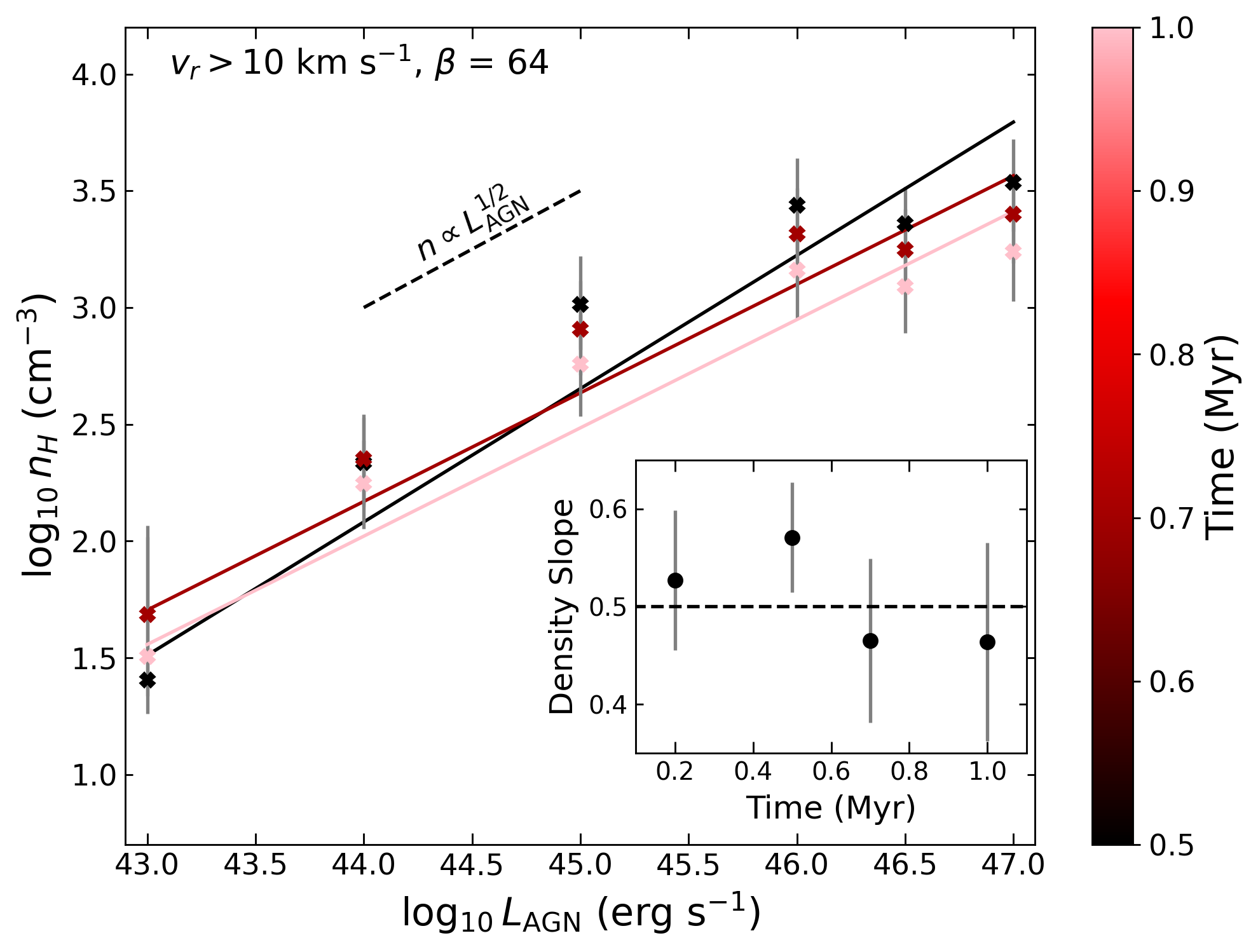}
    \caption[]{Main plot showing the density-AGN luminosity relationship at different times. The slope remains close to $0.5$ over the analyzed interval, as depicted in the inset panel. The normalization factor, however, varies with time but stabilizes after $t \approx 0.5$ Myr.}
    \label{fig:n-slope-time}
\end{figure}

The cumulative mass distribution does not exhibit a monotonic dependence on AGN luminosity. From Figure~\ref{fig:L-hist}, we see that both cloud density and size depend on $L_{\rm AGN}$. In our high-resolution simulations, the cloud size departs from a purely decreasing trend for $L_{\rm AGN} \geq 10^{45}$~erg~s$^{-1}$ -- showing the importance of resolution for estimating the clouds sizes. As a consequence, the CGC mass  does not follow a straightforward dependence on luminosity.

\subsubsection{Average Density Evolution}

In Figure \ref{fig:L-hist}, we showed the dependence of CGC density on AGN luminosity, $\lagn$, at $1$ Myr.
In Figure \ref{fig:n-slope-time}, we investigate whether this scaling relation holds at different simulation times. We confirm that between $0.2$ and $1$ Myr, the slope of the density-luminosity relationship remains fairly stable, oscillating between $\approx 0.4$ and $\approx 0.6$, as shown in the inset panel.

\begin{figure}
    \centering
    \includegraphics[width=\linewidth,trim={0cm 0cm 0cm 0cm},clip]{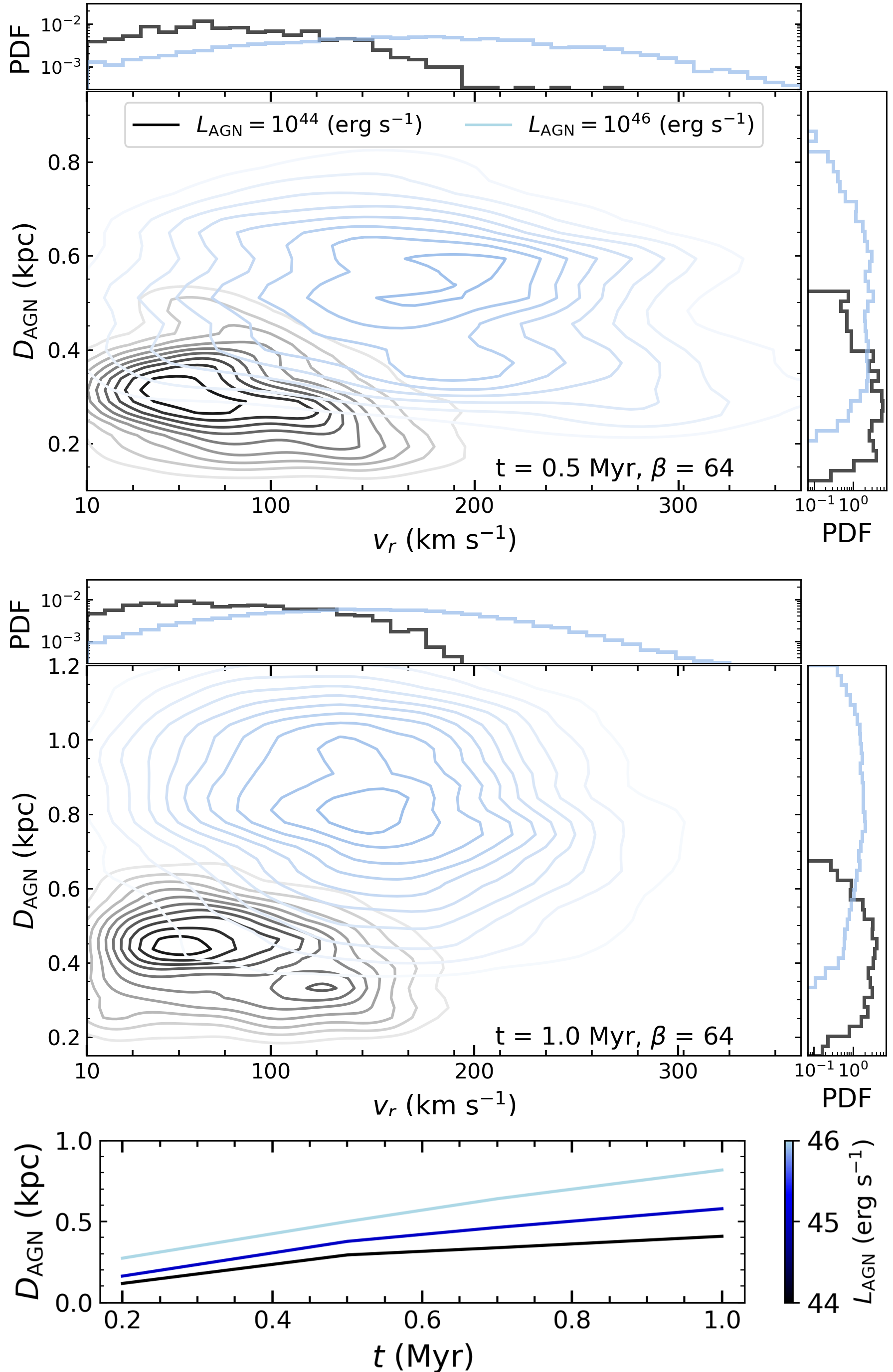}
    \caption[]{2D-diagram relating the radial velocity ($v_r$) and distance to the AGN ($D_{\rm AGN}$) for CGCs. Normalised histograms of $v_r$ and $D_{\rm AGN}$ are shown on the borders of the diagram. The red and blue lines represent simulations with $\lagn = 10^{44}$ erg s$^{-1}$ and $\lagn = 10^{46}$ erg s$^{-1}$, respectively; the upper and middle panels display data at $0.5$ Myr and $1$ Myr, respectively. This plot illustrates how CGCs populate the $v_r$-$D_{\rm AGN}$ space at different times and under different AGN luminosities, showing that more powerful AGNs tend to produce faster CGCs at greater distances. The lower panel shows the average $D_{\rm AGN}$ as function of time, it explicitly shows the positive relation between the AGN luminosity and the cloud location.}
    \label{fig:vD-hist}
\end{figure}

The consistency across time indicates that the relationship stabilizes rapidly, within timescales shorter than $200$ kyr, which is the time for the hot wind to cross the galaxy ($2$ kpc $/ \ 10^4$ km s$^{-1}$ $\approx 0.2$ Myr). However, the normalization factor, $n_0$, does appear to vary with time, $n_\mathrm{H}(\lagn, t) = n_0(t)(\lagn/10^{43})$ erg s$^{-1}$, likely as a consequence of the $R^{-2}$ scaling of the wind density \citep{Costa2020}.
$n_0$ falls until about $0.5$ Myr, after which it stabilises at $n_0(t > 0.5 \ \mathrm{Myr}) \approx 50$ cm$^{-3}$.

\subsubsection{Velocity and Spatial Distribution}

Figure \ref{fig:vD-hist} presents 2D histograms, showing CGCs' characteristic radial velocity ($v_r$) and radial distance from the galactic centre ($D_{\rm AGN}$), for simulations L44\_R64 (black contours) and L46\_R64 (light blue contours).
On the borders of each axis, 1D histograms give the distributions of each variable, normalised to unity. Figure \ref{fig:vD-hist} contains two panels, corresponding to $0.5$ Myr (upper panel) and $1$ Myr (lower panel).

The average CGC velocity increases with AGN power, keeping a broad distribution from $10$ km s$^{-1}$ to $300$ km s$^{-1}$ for all simulations. As already mentioned in Section \ref{subsec:Impact-Initial-ISM}, the fastest CGCs in our simulations, even those driven by $\lagn = 10^{47}$ erg s$^{-1}$, do not exceed $400$ km s$^{-1}$.

Meanwhile, the distribution of distances to the galactic centre broadens over time (note the radial distance from the AGN $D_{\rm AGN}$ is different from the upper and central panels). Indeed, the lower panel of Figure \ref{fig:vD-hist}, showing the average $D_{\rm AGN}$ increases with time and $\lagn$. This could occurs either because (1) CGCs originating near the centre move outwards at later times, or (2) faster-moving hot gas prompts the fragmentation of distant cool gas at larger radii.
Calculating the average velocity expansion as $\overline{v}_E \equiv \Delta D_{\rm AGN} / \Delta t$ using average distances and $\Delta t = 0.5$ Myr, we find $\overline{v}_E(\lagn = 10^{44}$ erg s$^{-1}$ $= 300$ km s$^{-1}$) and $\overline{v}_E(\lagn = 10^{46}$ erg s$^{-1}$) $= 600$ km s$^{-1}$. These velocities are twice the maximum observed velocities, indicating that the changes in $D_{\rm AGN}$ are not due to CGCs moving but rather the disruption of initial cool gas, at greater distances, by the faster hot gas. Hence, our detected CGCs are slow-moving; it is the fast-moving hot gas that forms new CGCs at larger $D_{\rm AGN}$, much before a CGC created at a closer distance to the galaxy centre could reach a farther location.

\subsection{Metal-line cooling}\label{subsec:MLC-Tfloor-Impact}

Our fiducial simulations adopt a temperature floor of $10^4$~K, as only primordial cooling is included in most runs. Our Sims2-Metal simulations probe metal-line cooling and a lower temperature floor of $10^2$~K (see Sims2-Metal in Table~\ref{tab:simulations-params}).

Including metal-line cooling enhances the cooling rate and allows for gas to cool down to below $10^4$~K.
We find that clouds become larger and more massive, but with somewhat lower mean densities (by $\sim 0.2$~dex compared to simulations with primordial cooling only; see Figure~\ref{fig:L-hist}), since most of the additional gas has $n_\mathrm{H} \lesssim 10^2$~cm$^{-3}$. Crucially, the qualitative trend $n_{\rm CGC} \propto L_{\rm AGN}^{1/2}$ remains unchanged, as it is primarily set by the AGN wind pressure; a different cooling prescription can affect the normalisation of this relation by $\sim 0.2-0.3$~dex.

In our simulations, changing the temperature floor from $10^4$~K to $10^2$~K had no appreciable impact on the CGC properties. Quantities such as CGC mass and density distributions are extremely similar for different temperature floors, and the global outflow structure is effectively identical between simulations that differ only in the choice of temperature floor.

\subsection{The relation between hot and cold outflow phases}\label{subsec:AGN-Gas-Impact}

\begin{figure}
    \centering
    \includegraphics[width=.96\linewidth,trim={0cm 0cm 0cm 0cm},clip]{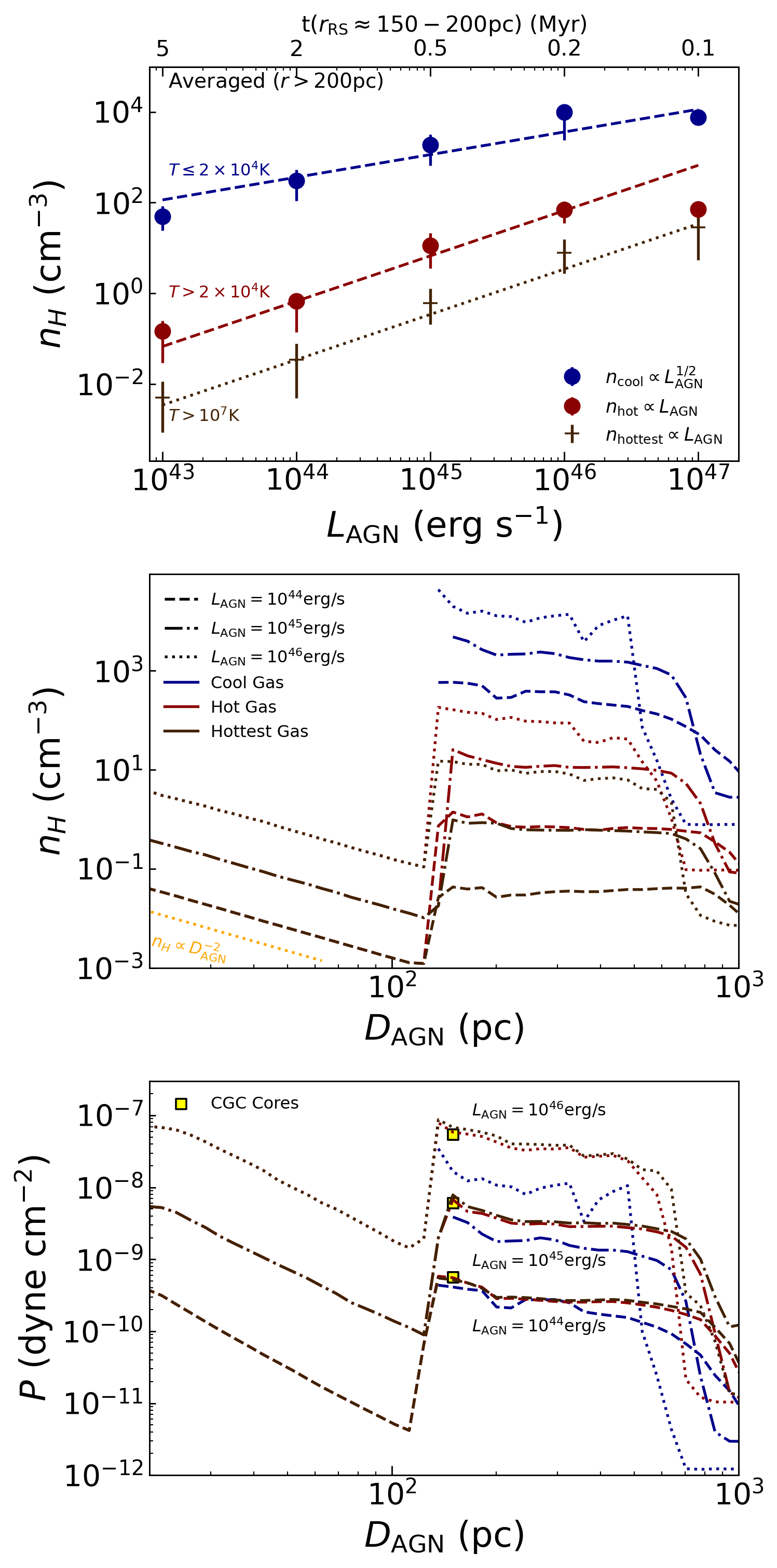}
    \caption[]{This figure includes all gas with $v_r > 10$ km s$^{-1}$, without applying the cloud selection process. The gas is divided into cool ($T \leq 2 \times 10^4$ K, blue), hot ($T > 2 \times 10^4$ K, dark red), and the ``hottest'' ($T > 10^7$ K, brown) --which is a subset of the hot phase-- phases. All simulations are analysed at the time when the reverse shock radius is approximately $r_{\rm RS} \approx 150$–$200$ pc, the time when each simulation was analysed is shown in the top x-axis of the upper panel.
    \textit{Upper panel:} Average hydrogen number density ($n_{\rm H}$) of each gas phase for $D_{\rm AGN} > r_{\rm RS}$. The cool phase line follows $\lagn^{s}$, with $s=0.5$, as we found from the cloud analysis (see Figure \ref{fig:L-hist}). The lines show a linear relation between $\lagn$ and $n_\mathrm{H}$ for the hottest phase and the hot phase up to $\lagn^{s} = 10^{46}$ erg s$^{-1}$.
    \textit{Middle panel:} Radial profile of hydrogen number density for the different phases. Line styles correspond to different AGN luminosities: $\lagn = 10^{44}$ erg s$^{-1}$ (dashed), $\lagn = 10^{45}$ erg s$^{-1}$ (dash-dotted), and $\lagn = 10^{46}$ erg s$^{-1}$ (dotted). A density jump is visible in the hot phase at $r_{\rm RS}$, coinciding with the appearance of the cool gas component.
    \textit{Lower panel:} Radial volume-weighted pressure profiles for the simulations. A corresponding pressure jump at $r_{\rm RS}$ is also observed. The plot reveals an increasing pressure gradient between the cool and hot phases with rising $\lagn$, indicating a more underpressured cool phase in the presence of more luminous AGN.
    }
    \label{fig:pressure-radius}
\end{figure}

We here compare the behaviour of cool and hot gas outflow phases. We no longer restrict our analysis to CGCs as in previous sections. We now consider all outflowing cool gas with $v_r \geq 10$ km s$^{-1}$ and $D_{\rm AGN} > r_{\rm RS}$, where $r_{\rm RS}$ is the reverse shock radius --
as the highly supersonic wind propagates, it drives a strong (forward) shock into the ISM \citep{Weaver1977, Zubovas2012, Faucher2012, Costa2014, Meenakshi2024}. In addition, a strong reverse shock decelerates and thermalises the fast wind (see, e.g. \citealt{Costa2014}, Figure 1).
This way, the analysis is irrespective of whether the gas is grouped into compact clouds or is more diffuse. Our aim is to test if the density - wind power relation presented in Section~\ref{subsec:Impact-LAGN} holds independently of how CGCs are defined.

In Figure \ref{fig:pressure-radius}, we analyse our different simulations from Sims1 at a fixed reverse shock radius $\approx 150-200$ pc.
The reverse shock radius is both time-dependent and sensitive to the wind power \citep{Costa2020}, which means that we here compare different simulations at different times.
Investigating the simulations at a fixed reverse shock radius ensures we compare them at a similar evolutionary stage.
Only the Sims1 suite is used here, as only these simulations are performed for a sufficiently long time for the reverse shock to reach a distance of $200-250$ pc in all simulations. The time $t(r_{\rm RS} \approx 150-200)$ for which the reverse shock is at radius $r_{\rm RS} \, \approx \, 150-200$ pc is shown in the upper panel of Figure \ref{fig:pressure-radius} for each $\lagn$.

In Figure \ref{fig:pressure-radius}, we divide the gas into three phases: the extensively-discussed cool phase ($T \leq 2 \times 10^4$ K), the hot phase ($T > 2 \times 10^4$ K), and a ``hottest'' phase ($T > 10^7$ K). The hottest phase is a subset of the hot phase, comprising the extremely hot, low-density end of the shocked wind and shocked ISM phases. At these high temperatures and low densities, we expect cooling times $\gg 10\, \rm Myr$, beyond our simulation run times.

The upper panel of Figure \ref{fig:pressure-radius} shows the hydrogen number density for these three phases, spatially averaged for $D_{\rm AGN} > r_{\rm RS}$. The cool phase follows a scaling of $\lagn^{s}$ with $s \approx 0.5$, consistent with the cloud-based analysis (see Figure \ref{fig:L-hist}). Restricting the analysis to $\lagn \leq 10^{46}$ erg s$^{-1}$, the relation becomes slightly steeper with $s \approx 0.75$.
The red (dashed) and brown (dotted) lines, for hot and hottest phases, respectively, show linear relations up to $\lagn = 10^{46}$ erg s$^{-1}$).
At $\lagn = 10^{47}$ erg s$^{-1}$, however, the trend breaks for cool and hot gas. At this high luminosity, the AGN wind appears to be so energetic that all cool gas is rapidly destroyed, and the resulting mixed-phase no longer cools rapidly enough.

The middle panel of Figure \ref{fig:pressure-radius} shows hydrogen number density radial profiles for simulations ranging from $\lagn = 10^{44}$ to $10^{46}$ erg s$^{-1}$. We show only three simulations to avoid overcrowding the panel. The reverse shock is visible as a density jump at $r_{\rm RS}$ for the hot phase. Cool, outflowing gas appears at radii $\gtrsim r_{\rm RS}$, but is absent in the freely-expanding wind region, where the the expanding, hot wind with $n_\mathrm{H} \propto D_{\rm AGN}^{-2}$ dominates the mass budget. Beyond the reverse shock, we can see that the density is highest for the cool phase. For this component, we also find that at fixed radius, the cool gas density scales approximately as $\propto L_{\rm AGN}$.
The number density falls with radius for every component, though much less steeply than $D_{\rm AGN}^{-2}$, very likely due to additional mass loading in the disc at these radii \citep{Ward2024} and additional phase mixing. The density decrease with radius affect the averaged density when we consider all gas, leading to a relation $n_\mathrm{H} \propto \lagn^s$, with $s < 1$.

The bottom panel of Figure \ref{fig:pressure-radius} shows radially-volume-weighted-averaged gas pressure for the same simulations.
For the hot gas phases, the pressure increases systematically with AGN luminosity, tracing the expected $\propto L_{\rm AGN}$ scaling (Section \ref{subsec:sims-data}).
For the cool gas phase, the pressure also increases with AGN luminosity, though with a sub-linear scaling.
While the pressure of the cool outflow component roughly matches that of the hotter phases, it appears lower for $L_{\rm AGN} \, = \, 10^{46} \, \rm erg \, s^{-1}$.
This deviation, however, disappears if we consider only high-density, cool outflowing, gas. The yellow squares shown in the bottom panel of Figure \ref{fig:pressure-radius} show the pressure of cool gas with the 10\% densest gas (check Figure \ref{fig:L-hist} for the density values) and $v_{\rm r} > 10$ km s$^{-1}$. We see these obey the same linear scaling as for hotter gas, remaining in approximate pressure equilibrium with these other phases. In Section~\ref{subsec:physical-interpretation}, we link this finding to the scaling relation between cool gas outflows and AGN luminosity found in this paper.

\section{Discussion}\label{sec:discussion}

In Section~\ref{subsec:physical-interpretation}, we present our physical interpretation of the relation we have found between outflowing cool cloud densities and AGN luminosity, shown in Figures \ref{fig:L-hist} and \ref{fig:pressure-radius}. 
Section \ref{subsec:implication-observations}, in turn, examines the implications of our results on estimates of AGN outflow masses, outflow rates and kinetic coupling efficiencies made in observations. We also explore their impact on observationally-derived outflow scaling relations.  
Finally, in Section~\ref{subsec:limits}, we address the limitations of our current simulations and suggest future improvements.

\subsection{Physical Interpretation}\label{subsec:physical-interpretation}

\subsubsection{The origin of the Density-Power relation}\label{subsubsec:density-interpretation}
 
Figures \ref{fig:L-hist}, \ref{fig:n-slope-time}, and \ref{fig:pressure-radius} reveals a clear relation between CGC density and wind kinetic power.  
If we consider only high-density CGC cores, a density-power relation holds across the entire range of $\lagn$, from $10^{43}$ erg $s^{-1}$ to $10^{47}$ erg $s^{-1}$.

The spatial association of CGCs with gas tails behind ISM clumps, the dependence of their formation on radiative cooling (see \textit{W24}), and the power law cloud mass distributions seen in Sections~\ref{subsec:Impact-Initial-ISM} and~\ref{subsec:temporal-evolution} are all strong indications that CGCs form via rapid cooling in mixing layers composed of ISM gas and hot wind material, as outlined in \citet{Gronke2018}. 

As shown in the bottom panel of Figure~\ref{fig:pressure-radius}, the pressure of the cool, outflowing phase scales approximate linearly with the pressure of the surrounding hot outflow phase.
The link between cool gas outflows and AGN luminosity found in this paper is thus driven by the tendency of dense, cool clouds in the outflow to settle into pressure equilibrium with the hot, outflowing medium. 
Thus

\begin{enumerate}
    \item[\textbf{i.}] the AGN wind impinges on initial cool clumps, producing a tail of mixed wind and ISM fluid,
    \item[\textbf{ii.}] cool clouds condensing out of this turbulent wake are confined by a higher pressure for higher AGN luminosity and wind power,
    \item[\textbf{iii.}] At a fixed temperature of $T \sim 10^4 \, \rm K$, cool gas needs to reach higher density in order to settle into pressure equilibrium with its surrounding medium. Since the pressure of the hot wind component scales with wind power and AGN luminosity, the density of surviving, cool gas clouds also scales with $L_{\rm AGN}$. 
\end{enumerate}

Our result does rest on the assumption that the hot gas pressure scales with AGN luminosity. This follows from assuming a fixed AGN wind velocity at injection and a fixed momentum loading factor such that the wind mass injection rate scales as $\propto L_{\rm AGN} / (v_{\rm wind} c)$. The $v_{\rm wind} \propto L_{\rm AGN}^{0.5}$ scaling found by \citet{Gofford2015} may imply a shallower scaling than predicted in this paper. These scalings are, however, still uncertain. For instance, \citet{Matzeu2023} report a shallower relation, with $v_{\rm wind} \propto L_{\rm AGN}^{0.19}$, not too different from our model assumptions.  

While the exact scaling between cool gas outflow density and AGN luminosity may thus depend on how the latter relates to small-scale wind properties, it only arises thanks to the presence of a large-scale hot medium that retains its pressure. 
The existence of such a phase is the hallmark of an ``energy-driven'' outflow. As discussed at length in various studies \citep{Zubovas2014, Nims2015, Costa2014, Costa2020}, this hot medium has a remarkably low emissivity.
The observational detection of a link between the density of cool outflows in system and AGN luminosity indirectly thus provides evidence for the existence of such hot, ``energy-driven'' bubbles.

\subsubsection{Cloud velocities}\label{subsubsec:velocity-interpretation}

Across the entire simulation set, we do not observe CGCs with velocities higher than $v_r \gtrsim 500 \, \rm km \, s^{-1}$ (Figure \ref{fig:vD-hist}). 
In the cooling-mediated cloud entrainment scenario outlined in \citet{Gronke2018}, cool clouds become entrained on a drag timescale $\sim \left( \rho_{\rm cl} / \rho_{\rm w}\right) \left( r_{\rm cl} / v_{\rm w}\right)$. 
Given the presence of efficient cooling in our simulations, these CGCs could be accelerated to high velocities prior to being crushed if we ran the simulation for a time larger than the acceleration timescale.

A detailed calculation of the acceleration timescale, $t_{\rm acc}$, is provided in Appendix \ref{app:acceleration-timescales}. For our parameters, 

\begin{equation}
    t_{\rm acc} \approx 25 L_{45}^{-2/3}D_{100}^2v_4^{-1} \, \rm Myr \, , 
    \label{eq:tacc-myr}
\end{equation}

\noindent where $L_{45} = \lagn / 10^{45} \text{ erg s}^{-1}$, $D_{100} = D_{\rm AGN}/ 100 \text{ pc}$, and $v_4 = v_{\rm wind} / 10^4 \text{ km s}^{-1}$. Considering CGCs are typically further than $100$ pc from the galaxy centre (as shown in Figure \ref{fig:vD-hist}), the acceleration timescales for winds with injection velocities equal to $10^4$ km s$^{-1}$ and $5000$ km s$^{-1}$ are above $20$ Myr for $\lagn = 10^{45}$ erg\,s$^{-1}$, and this number is even larger for winds that slowed down. The calculated $t_{\rm acc}$ greatly exceeds our simulation duration and also exceeds the expected AGN activity time.  

According to Equation \eqref{eq:tacc-myr}, achieving high-velocity cool gas would require substantially longer simulation run-times. However, individual AGN high-accretion episodes are very unlikely to last tens or hundreds of Myr. The absence of CGCs with $v_r \gtrsim 500$ km s$^{-1}$ in our simulations is thus more likely attributed to missing physical mechanisms, such as more efficient cooling (e.g. metal line cooling) and self-gravity to form ultra-dense outflowing cool gas. Based on their estimated Jeans mass (For $T = 20000$ K and $n_{\rm H} \sim 10^4$ cm$^{-3}$, $M_{\rm Jeans} \approx 10^6 \ \msun$), our CGCs are not expected to be self-gravitating, but the presence of self-gravity may increase the cloud's survival times \citep{Mandal2024}. Indeed, in new simulations featuring metal-line cooling, otherwise adopting the same setup as in \textit{W24}, cool gas is found with speeds $\gtrsim 600 \, \rm km \, s^{-1}$ (Ward et al., Haidar et al., in prep.).
From Equation~\ref{eq:tacc-myr}, it also follows that cloudlets located at shorter distances to the AGN should become entrained on shorter timescales. Another possible route to faster cool gas clouds is thus to consider initial clumps at even smaller distances to the AGN than considered in this paper. 
Finally we note that our initial conditions comprise a static gas distribution. The interaction of an AGN wind with cool gas which is already moving fast \citep[e.g.][]{Costa2015} is likely to result in faster cool outflows than seen here.

\subsection{Implication for observations of AGN outflows}\label{subsec:implication-observations}

We here explore the potential implications of our theoretical results on observational measurements of AGN-driven outflows. We first look at mass outflow rates and kinetic powers (Section~\ref{subsec:impact-outflow-rates}), before focusing on the implications on scaling relations between mass outflow rate and AGN luminosity (Section~\ref{subsec:scaling-relations}). Finally, we compare our predicted relation between AGN luminosity and outflow density with available measurements in the literature (Section ~\ref{subsec:obs-comparison}).  

\subsubsection{Outflow rates from cloud densities}\label{subsec:impact-outflow-rates}

\textit{W24} provide describes our methodology to extract outflow properties from the simulations, in a way that is consistent with observational approaches. It is common to combine outflow mass, $M_{\rm OF}$ (see Section~\ref{subsec:scaling-relations}), outflow velocity, $v_{\rm OF}$, and an outflow radius, $R_{\rm OF}$, to estimate a time-averaged mass outflow rate $\dot{M}_{\rm OF}$, and a kinetic power $\dot{E}_k$, following
\begin{equation}
\begin{split}
    \dot{M}_{\rm OF} = B \cdot \frac{M_{\rm OF} \cdot v_{\rm OF}}{R_{\rm OF}}, \, {\rm and} \\
    \dot{E}_k = \frac{1}{2} \cdot \dot{M}_{\rm OF} \cdot v_{\rm OF}^2.
\end{split} 
\label{eq:mdot-outflow}
\end{equation}
Here $B \sim 1$ depends on the assumed geometry \citep{Gonzalez-Alfonso2017,Harrison2018}. 
The following analysis makes use of {\em all gas} in our simulations with radial velocities $v_r > 10$ km s$^{-1}$, regardless of whether it is located within a CGC or not, noting that Figure~\ref{fig:pressure-radius} confirms that both methods yield the same density--wind power relation. Time-averaged outflow properties are computed by summing over all cells with $v_{\rm r} > 10 km s^{-1}$ and dividing by the time passed since the AGN wind was initiated. 
\textit{W24} discuss in detail the challenges in defining robust outflow velocities and radii from observations to use in these equations. Here, we primarily focus on the implications of tracing a limited range of gas densities in the outflows to infer mass outflow rates and kinetic powers. 

\begin{figure}
    \centering
    \includegraphics[width=\linewidth,trim={0cm 0cm 0cm 0cm},clip]{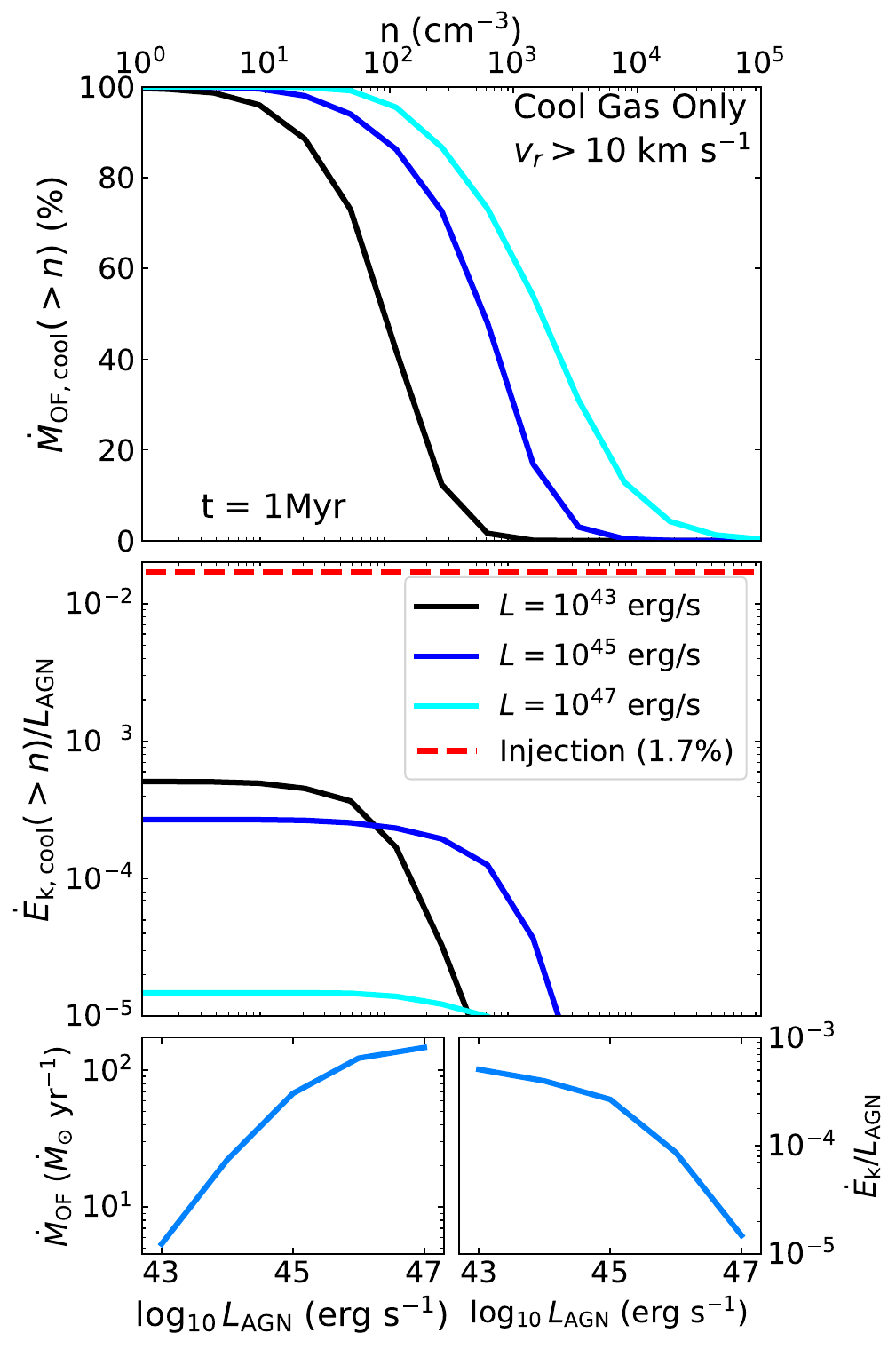}
    \caption{\textit{Top Panel:} Fraction of the mass outflow rate for gases exceeding certain density thresholds. For instance, gas with densities above $10^3$\,cm$^{-3}$ accounts for approximately 30\% of the total outflowing CGC mass at $\lagn = 10^{45}$ erg\,s$^{-1}$ and about 60\% at $\lagn = 10^{47}$ erg\,s$^{-1}$. \textit{Middle Panel:} Kinetic coupling energy of the outflowing CGCs, where the ratio $\dot{E}_k / \lagn$ is low, indicating that this gas phase does not carry a significant amount of the total energy. \textit{Bottom Panels:} $\dot{M}_{\rm OF}$ and $\dot{E}_k/\lagn$ as function of AGN luminosity for the cool phase.}
    \label{fig:mdot_ek}
\end{figure}

In the bottom left panel of Figure~\ref{fig:mdot_ek}, we show $\dot{M}_{\rm OF}$ for cool gas as a function of AGN luminosity. We reproduce the result from \textit{W24} that cool mass outflow rate increases with increasing luminosity. However, our new result shows that the distribution of cool, outflowing mass across density is luminosity-dependent (Section \ref{subsec:Impact-LAGN}). In the top panel of Figure~\ref{fig:mdot_ek} we present the cumulative distribution of $\dot{M}_{\rm OF}$ as a function of gas density. This highlights a significant fraction of the total outflow rates could be missed when the observations are not sensitive to the lowest densities, with this effect being most severe for the lowest AGN luminosities. 
For example, almost 100\% of the mass outflow rate is associated with gas with $n_{\rm H} > 50$\,cm$^{-3}$ for the most luminous AGN with $L_{\rm AGN}=10^{47}$\,erg\,s$^{-1}$. However, for $L_{\rm AGN}=10^{43}$\,erg\,s$^{-1}$, $\approx $30\% of the mass outflow rate has $n_{\rm H} > 50$\,cm$^{-3}$. Conversely, the densest gas, with $n_{\rm H} \geq 10^3$\,cm$^{-3}$, has negligible contribution (i.e., $\sim0.6$\%) to the cool outflowing mass rate for low luminosity AGN ($\lagn = 10^{43}$ erg\,s$^{-1}$), but constitutes $\approx 60 \%$ of the cool outflow mass rate for the most luminous AGN ($\lagn = 10^{47}$ erg\,s$^{-1}$).

\begin{figure*}
    \centering
    \includegraphics[width=\linewidth]{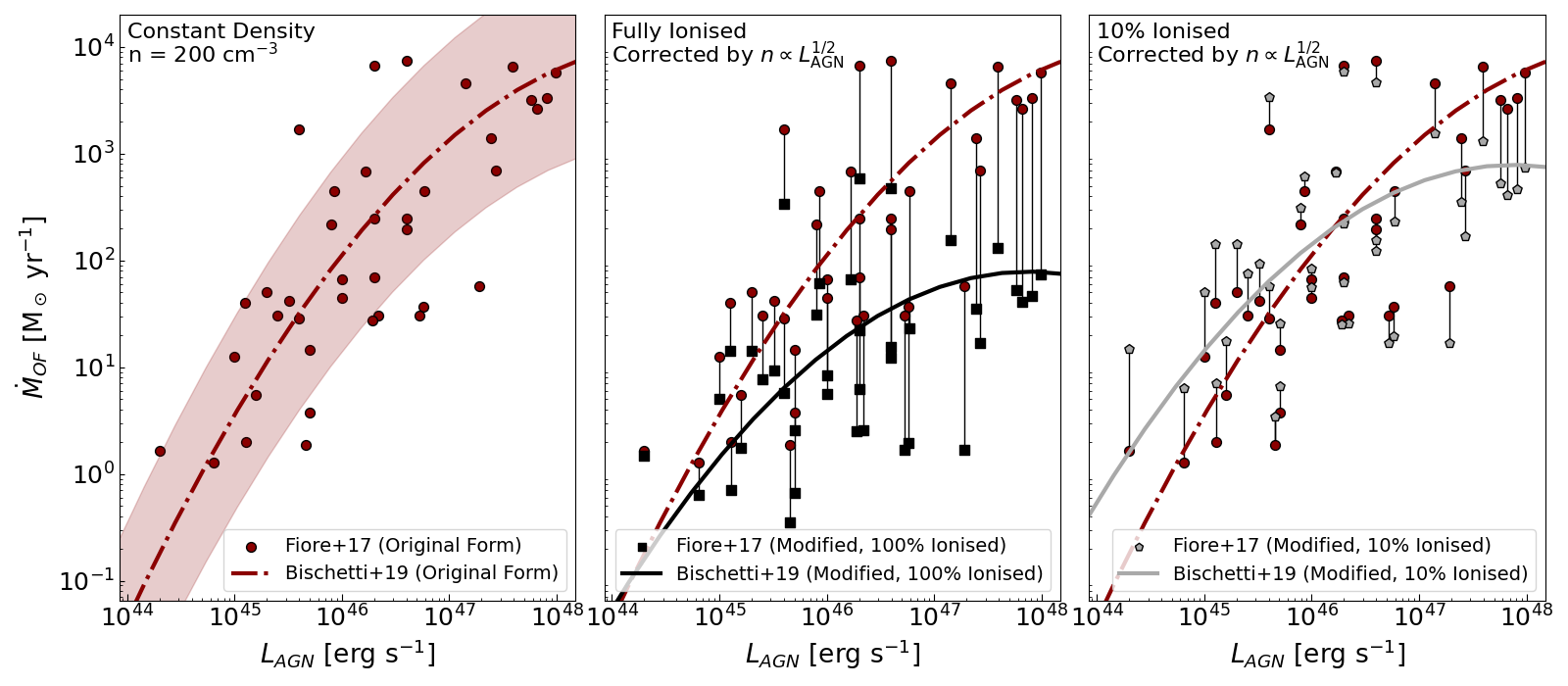}
    \caption{Comparative analysis of observed AGN mass outflow rates, $\dot{M}_{\rm OF}$, before and after accounting for our simulation-inferred dependency between $\lagn$ and density. Left panel: The original $\dot{M}_{\rm OF}$ values for individual systems reported by \protect\cite{Fiore2017} (data points) and the observed relation for these data, and a wider compilation, derived by \protect\cite{Bischetti2019} (curve and shaded region). The curve and data points all assume a constant electron density across all sources. Center and right panels: Recalculated $\dot{M}_{\rm OF}$ values using our density-luminosity relation for different total density to electron density ratios of $\epsilon_\mathrm{e} = 1$ (fully ionised) and $0.1$ (partially ionised), respectively. The directions and magnitudes of the modifications to the original data points are highlighted by the lines linking the original (red circles) and the modified values (black squares or grey pentagons). The recalculated relation is shown by the black and grey solid curves. Assuming $\lagn$-dependent electron density values, instead of a constant values across all targets, has the effect of flattening the relation and significantly reducing the outflow rates for the sources with highest $\lagn$.}

    \label{fig:fiore-comparison}
\end{figure*}

In the bottom right panel of Figure~\ref{fig:mdot_ek} we show $\dot{E}_k$/$L_{\rm AGN}$ for cool gas as a function of AGN luminosity. We have normalised the kinetic power by AGN luminosity to give the ``kinetic coupling efficiency'' often quoted in the literature.
However, we note that this phase-dependent quantity should not be confused with the AGN feedback coupling efficiency parameter adopted in numerical simulations (see discussion in \citealt{Harrison2018, Ward2024}). 
We here reproduce the result from \textit{W24} that the kinetic coupling efficiency decreases with increasing AGN luminosity for the cool gas phase, though the exact form of this relation is sensitive to the velocity used to define outflow (see \textit{W24}). 

It is important to note that while the initial injected kinetic power of the wind is 1.7\% of $L_{\rm AGN}$ in our simulations with a wind speed of $10^4 \, \rm km \, s^{-1}$ (see Section \ref{subsec:sims-data}), only a tiny fraction is carried by the cool outflows, amounting to a negligible contribution for the highest luminosities (\textit{W24}; Figure~\ref{fig:mdot_ek}). In this work, we find that inferring very low kinetic coupling efficiencies is even more likely if the observations are only sensitive to a limited range of densities, and the level of this effect is luminosity-dependent. This is highlighted in the middle panel of Figure~\ref{fig:mdot_ek}, where we show the cumulative distribution of kinetic coupling efficiency as a function of gas density for three different bins of AGN luminosity. For example, for AGN luminosities of $L_{\rm AGN}=10^{45}$erg\,s$^{-1}$ and $L_{\rm AGN}=10^{47}$erg\,s$^{-1}$ the total kinetic coupling efficiencies in the cool phase are 0.05\% and 0.0015\%, respectively. However, this drops even further ($\lesssim$0.001\%) when only considering high-density gas, with $n_{\rm H} \gtrsim 10^{3}$ cm$^{-3}$, showing that very low kinetic coupling efficiencies could be inferred from tracers of cool, dense gas. Taken at face value, these low kinetic coupling efficiencies could be interpreted as indicating a weak impact of the outflow on the galaxy. However, it is clear that the AGN outflows in our simulations dramatically impact the ISM (see Figure \ref{fig:maps}); the majority of the energy is simply carried in hotter and lower density gas phases. 

\subsubsection{Implications for observed outflow scaling relations}\label{subsec:scaling-relations}

With the broad goal of assessing the ability of AGN outflows to impact their host galaxies, it is common for observationally-inferred mass outflow rates to be compared to AGN luminosity, to produce outflow scaling relations (e.g., \citealt{Fiore2017,Harrison2018,Bischetti2019,Musiimenta2023,Bertola2025}). These can be constructed using a variety of absorption lines and emission lines to trace various phases of outflowing gas (\citealt{Harrison2024}). Due to the brightness of the line, and the abundance of available (rest frame) optical spectroscopy, the [O~{\sc iii}]$\lambda$5007 emission line is the most common tracer of the outflow kinematics of ionised gas for AGN. Whilst [O~{\sc iii}] may exist in gas up to $\sim$10$^{6}$\,K (\citealt{Katz2022}), temperature sensitive diagnostics from observations of AGN emission-line regions infer that the typical temperatures of the [O~{\sc iii}]$\lambda$5007 emitting gas to be $\approx$(1--2)$\times$10$^{4}$\,K, (\citealt{Tadhunter1989,Binette1996,Wilson1997,Perna2017,Revalski2018b}). Therefore, the temperature of this gas meets the definition of ``cool gas'' in our simulations and we choose to focus on relevant outflow scaling relations that use [O~{\sc iii}], as the primary outflow kinetic tracer, with H$\alpha$, H$\beta$, or [O~{\sc iii}], as the primary outflow mass tracer. Nonetheless, our qualitative results are expected to  be applicable to outflow tracers of colder gas phases, such as molecular emission lines (e.g., \citealt{Fluetsch2019,RamosAlmeida2022}). 

Based on an observed emission-line luminosity of outflowing material, $L_{\rm el, out}$, outflowing gas masses are obtained for the observed outflows following
\begin{equation}
    M_{\rm OF} = C \cdot \frac{L_{\rm el, out}}{n_\mathrm{e}}.
    \label{eq:mass-outflow}
\end{equation}
\noindent In this context, the constant $C$ is a value that depends on the emission line being used (e.g., H$\beta$, H$\alpha$), and $n_{\rm e}$ is the electron density. This mass can then be substituted into Equation~\eqref{eq:mdot-outflow}, to infer mass outflow rates and kinetic luminosities. 

We present a published scaling relation which makes use of optical emission line measurements, in the left panel of Figure~\ref{fig:fiore-comparison}. This shows mass outflow rate versus AGN luminosity for warm ionised outflows from the compilation in \cite{Fiore2017} and the derived relation from \cite{Bischetti2019} (who performed a fit to the combined data from \citealt{Fiore2017} and \citealt{Fluetsch2019}).  Crucially, for these results a constant electron density $n_\mathrm{e} = 200$ cm$^{-3}$ is assumed in the calculation of mass outflow rate (see Appendix A and B in \citealt{Fiore2017}) due to the lack of density-sensitive diagnostics across the combined samples (see also \citealt{Musiimenta2023,Bertola2025}). 

If outflow densities scale as $n_{\rm H} \propto \lagn^{-1/2}$ (Section \ref{subsec:Impact-LAGN}; Figures \ref{fig:L-hist} and \ref{fig:pressure-radius}), assuming a constant density in Equation~\eqref{eq:mass-outflow} is expected to lead to underestimating (overestimating) the total mass outflow rate for low-power (high-power) AGN. 
Assuming that the electron density $n_\mathrm{e} = \epsilon_\mathrm{e} 1.24 n_{\rm H}$ --for a fraction of 76\% of hydrogen--, we can use Equation \eqref{eq:mass-outflow} to rewrite Equation \eqref{eq:mdot-outflow} as $\dot{M}_{\rm OF} = \kappa n_\mathrm{e}^{-1}$, with $\kappa =  BC \ L_{\rm el, out} \ v_{\rm OF} R_{\rm OF}^{-1}$. Using our derived relation between hydrogen number density $n_{\rm H}$ and $\lagn$ as $n_{\rm H} = 50 (\lagn/10^{43} \text{erg s}^{-1})^{1/2}$,
we can thus derive a modified mass outflow rate, $\dot{M}_{\rm OF}^{\rm Mod} = \mathcal{Z} \dot{M}_{\rm OF}$ as
 \begin{equation}
     \mathcal{Z} = \frac{\dot{M}_{\rm OF}^{\rm Mod}}{\dot{M}_{\rm OF}} = \frac{\kappa n_\mathrm{e}^{-1}}{\kappa / 200 \text{ cm}^{-3}} = \frac{200 \text{ cm}^{-3}}{\epsilon_\mathrm{e} n} = \frac{5}{\epsilon_\mathrm{e}}\left( \frac{\lagn}{10^{43} \text{erg s}^{-1}} \right)^{-\frac{1}{2}}.
\label{eq:mof-correction}
 \end{equation}
This equation shows the expected correction to the derived mass outflow rates (with a fixed assumed density) to include the luminosity-dependent density.

In the middle and right panels of Figure \ref{fig:fiore-comparison}, we display modified values of mass outflow rate measurements from \cite{Fiore2017}, and a modification to the relation from \cite{Bischetti2019}, following Equation~\eqref{eq:mof-correction}. The middle and right panels assume $\epsilon_\mathrm{e} = 1$, as the fully ionised case, and $0.1$ as a low-ionisation case, respectively. This Figure illustrates how incorporating a luminosity dependence of outflow densities modifies the relation between the mass outflow rate, $\dot{M}_{\rm OF}$, and AGN luminosity, compared to assuming constant density across all luminosities. For luminosities below $\sim 2.5 \epsilon_\mathrm{e}^{-2} \times 10^{44}$ erg\,s$^{-1}$ (or $n_e < 200\epsilon_e$cm$^{-3}$), we infer an increase to correct $\dot{M}_{\rm OF}$, with a decrease beyond this luminosity. 

In summary, we find that neglecting a luminosity-dependent variation in outflow density can lead to errors of up to three orders of magnitude in the calculated outflow rates (Figure~\ref{fig:fiore-comparison}). The overall impact of a luminosity-dependent outflow density, compared to a fixed assumed density for all sources, is to flatten the relation between mass outflow rate and AGN luminosity.\footnote{In Figure \eqref{fig:fiore-comparison}, we note that we have assumed the unrealistic situation that the ionisation fraction, $\epsilon_e$, does not change with density nor AGN luminosity. However, if we recalibrate the correction for $\dot{M}_{\rm OF}$ to include an increasing ionised gas fraction with increasing $\lagn$, this relation will flatten even further.} This experiment shows the importance of accurately measuring density to estimate the mass outflow rates not only for the absolute values (Section~\ref{subsec:obs-comparison}), but also for the {\em form} of the scaling relation itself. 

\subsubsection{A comparison to electron densities from observations}\label{subsec:obs-comparison}

When density-sensitive diagnostics have been obtained for observations of warm ionised outflows, much evidence suggests that the outflowing gas has higher electron densities than typically found for quiescent gas (\citealt{Perna2017,ForsterSchreiber2019,Mingozzi2019,Holden2023}), and higher than typically assumed $\sim$100--200\,cm$^{-3}$ as a characteristic density in several studies  (Section~\ref{subsec:scaling-relations}; \citealt{Genzel2014,Fiore2017,Bischetti2019}). This may be in broad agreement with our prediction that luminous AGN enhance the densities in outflowing clouds. However, we now investigate if this can be quantitatively tested with available density measurements in the literature.

From an observational perspective, there are multiple methods for measuring electron densities using the emission lines in optical spectroscopy. \cite{Davies2020} discuss and compare three of the most common approaches: (1) using the density sensitive [S~{\sc ii}] $\lambda\lambda$ 6716, 6731 \AA\ doublet ratio (see e.g., \citealt{Osterbrock2006}); (2) combined ratios of strong and transauroral lines of [S~{\sc ii}] with strong and auroral lines of [O~{\sc ii}] (see e.g., \citealt{Holt2011}); and (3) assumptions based on the ionisation parameter $U$, derived from the strong line ratios [N~{\sc ii}] /H$_\alpha$ and [O~{\sc iii}] /H$_\beta$ (see e.g., \citealt{Baron2019}). The first of these methods is only sensitive up to electron densities in the range $n_{e}\sim50-5000$\,cm$^{-3}$ (e.g., \citealt{Osterbrock2006,Revalski2018a}), whilst the latter two methods are sensitive to much higher density gas (noting they also rely on more complex assumptions, such as the ionising continuum and the metallicity). Indeed, for a sample of 11 galaxies, with AGN luminosities ranging from $10^{43}$ to $10^{44.5}$ erg\,s$^{-1}$, \cite{Davies2020} find lower electron densities for the first method; with average values of $\log_{10} (n_e$/cm$^{-3}$)\,= 2.54, 3.28, and 3.68, across the three methods, respectively.  This re-highlights the importance of understanding the completeness of the observational tracers to different densities of outflowing gas (Section~\ref{subsec:impact-outflow-rates}). 

\begin{figure}
    \centering
    \includegraphics[width=\linewidth,trim={0cm 0cm 0cm 0cm},clip]{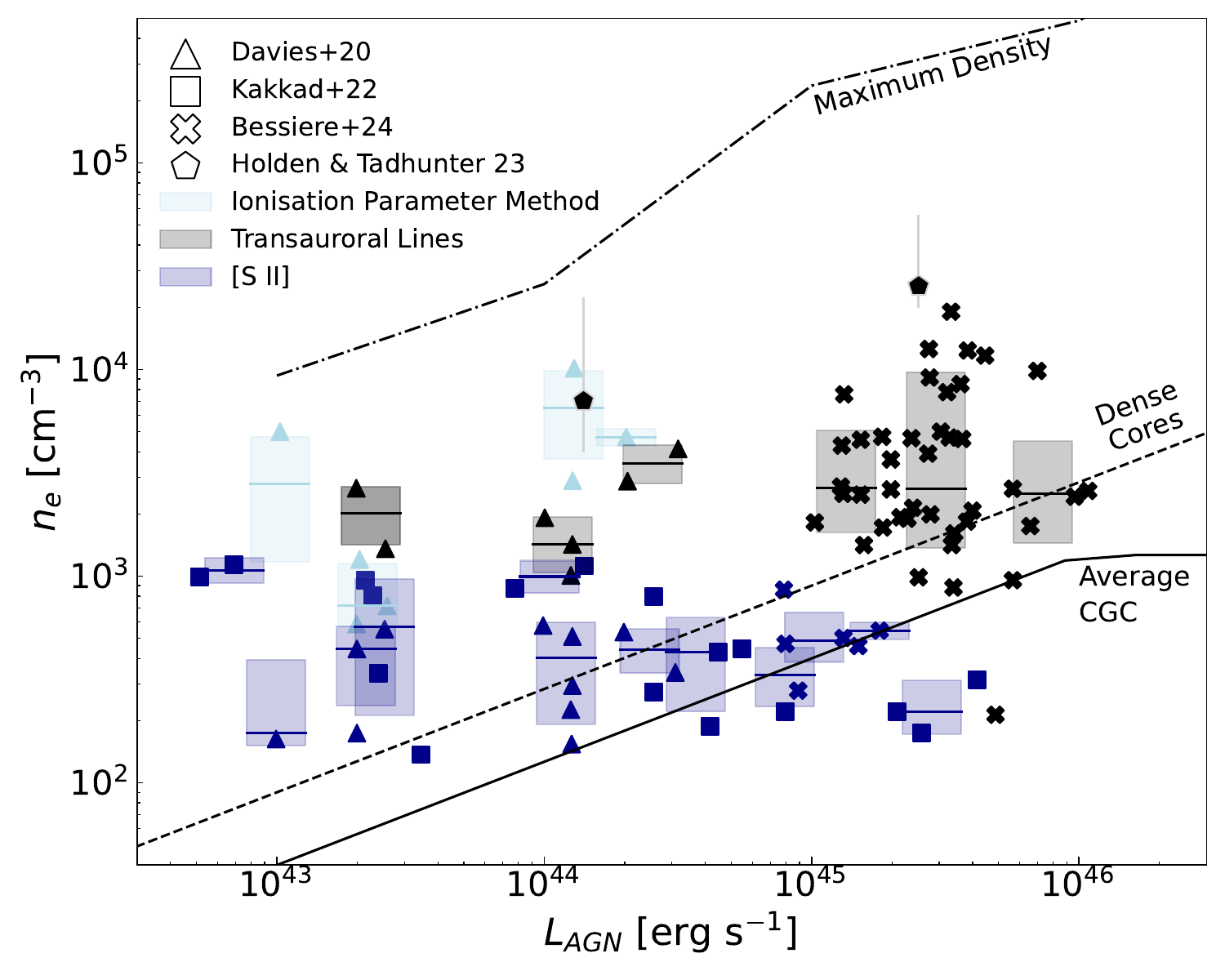}
    \caption{Diagram relating electron density and AGN luminosity from various studies: crosses from \protect\cite{Bessiere2024}, squares from \protect\cite{Kakkad2022}, triangles from \protect\cite{Davies2020}, pentagons from \protect\cite{Holden2023}, with different colours indicating the methods used to calculate $n_e$ ([S~{\sc ii}]  doublet, transauroral and auroral lines, and ionization parameter), shown in navy, black, and light blue respectively. The coloured regions, representing mean values for each dataset for a range of luminosities (0.5 dex bins). The lines are our simulations results: the solid line represents the average CGC density; the dashed line represents the average of the high density CGC cores; and the dot-dashed line indicates the maximum densities found inside the CGCs.
}
    \label{fig:bessiere-comparing}
\end{figure}

In Figure \ref{fig:bessiere-comparing} we show electron density measurements from observations of warm ionised outflows, as a function of AGN luminosity for the \citealt{Davies2020} sample, in addition to samples from \cite{Kakkad2022}, \cite{Holden2023} and \cite{Bessiere2024}, which also make use of one, or more, of the three methods described above for obtaining electron densities\footnote{We note that this is not an exhaustive comparison, and other work has attempted to make individual measurements of electron densities in AGN outflows, albeit typically focussed on the [S~{\sc ii}] doublet method (e.g., \citealt{Perna2017,Kakkad2018,Baron2019})}. Coloured regions represent the mean values and, respective uncertainties, for the measurements using the same method in 0.5\,dex luminosity bins. 

The [S~{\sc ii}] measurements show the expected lower electron densities than the other two methods; as this tracer is insensitive to the highest density gas. Furthermore, as discussed in \cite{Davies2020}, whereas the other tracers are dominated by fully ionised regions of clouds, the [S~{\sc ii}] doublet is tracing gas that can be partially ionised. This means that the actual hydrogen densities might be significantly larger than the inferred electron densities from [S~{\sc ii}] and may not valid for estimating the densities of [O~{\sc iii}] emitting gas. However, it is also important to note this [O~{\sc iii}] emitting gas is only tracing a small subset of the total outflowing gas, which is fully ionised, and typically high density. Based on our simulations, if the outflow tracer we are using is only relevant for a fraction of the total clouds, such as high density cores, a significant fraction of the total outflowing mass and energy could be missed (Section~\ref{subsec:impact-outflow-rates}; Figure~\ref{fig:mdot_ek}). Furthermore, spatially-resolved measurements show a wide variation of densities within the same galaxy, which might reach very low values in the more extended regions (e.g., \citealt{Venturi2023}).

Unfortunately the compilation of observations presented in Figure~\ref{fig:bessiere-comparing}, can not easily be used to compare directly to our simulation results. This is because of the inhomogeneous nature of the observations, and a lack of constraints on the mass-weighted average densities of the cool outflows (e.g., by carefully combining a range of diagnostics in the same samples). This should constitute future work. Nonetheless, to assess a potential correlation between electron density and AGN luminosity, we perform three standard statistical tests. The Pearson correlation coefficient yields $r \approx 0.34$, while the Spearman and Kendall rank correlation tests give $\rho \approx 0.48$ and $\tau \approx 0.34$, respectively. These values suggest a mild positive correlation, particularly in the rank-based tests, but the relationship is not strong. While not conclusive, this preliminary analysis indicates there may be a trend worth exploring further with a more homogeneous samples and complete methods.

We compare the observations to our simulation results as lines (assuming $n_\mathrm{e} \sim 1.24n_\mathrm{H}$, for a hydrogen fraction of 76\%) in Figure~\ref{fig:bessiere-comparing}. The solid black line represents the average density value across all CGCs and the dashed line indicates the average core density (see Section \ref{subsec:find-clouds}). The dash-dotted line in Figure \ref{fig:bessiere-comparing} refers to the maximum density value found inside a CGC. When considering the denser core regions of the CGCs, the average density values become comparable to observational values inferred from (trans)auroral emission line methods and the ionisation parameter method. This trend is consistent with the idea that these methods trace the higher density (fully ionised) regions of outflowing clouds. Even so, the average densities inferred from our simulated cores (dashed line in Figure~\ref{fig:bessiere-comparing}) are typically on the lower end of the observationally-derived values, particularly for the lower luminosity systems. This may be partly due to the fact the observational diagnostics, are not very sensitive to the very lowest densities ($\lesssim$50\,cm$^{-3}$), which we predict to be increasingly important with decreasing luminosity (Section~\ref{subsec:impact-outflow-rates}; Figure~\ref{fig:mdot_ek}). Future work could involve using the simulations to predict specific emission lines, and enable a more direct comparison to specific observational diagnostics (see Section~\ref{subsec:limits}). 

\subsection{Cloud crushing and AGN winds}\label{subsec:cloud-crushing-sims}

Our numerical experiments follow the interaction between a dense initial ISM cloud and a fast, high specific energy AGN wind. We find a population of compressed outflowing, cool clouds on galactic scales ($\approx 0.5-2$~kpc from the galaxy's centre), with properties in line with those predicted by smaller-scale, cloud-crushing simulations (e.g. \citealt{BandaBarragan2019,Gronke2020}).

The cloud-crushing literature is extensive and covers a wide range of astrophysical contexts (stellar winds, AGN winds, jets). Several studies explore this process in the context of AGN winds (e.g. \citealt{Mellema2002, Zubovas2014b, Bourne2014, Dugan2017, Mandal2024, Lauvzikas2024, Zubovas2024}). 
For example, \citet{Dugan2017} investigate the small-scale impact of AGN feedback on an individual cloud, exploring different wind parameters such as velocity and density, which are directly related to $L_{\rm AGN}$. In their simulations, the wind impacts a large initial clump, followed by simultaneous stripping and collapse into smaller clouds. They find that ram pressure is the key parameter determining the cloud response. In our simulations, the wind ram pressure, which is proportional to $L_{\rm AGN}$, appears to be the dominant mechanism for clearing gas from the nuclear region. However, once the wind shocks, the resulting hot gas provides substantial thermal pressure that confines cold cloudlets entrained in the outflow.
More recently, \citet{Lauvzikas2024} investigate the interaction of AGN winds with isolated turbulent clouds, finding widespread compression and fragmentation across AGN wind properties. Probing different wind densities, lower velocities, and varying temperatures (parameters plausibly linked to the distance from the AGN and galactic environment), they show that turbulence-driven overdensities can assemble into massive clouds ($10^2 \-- 10^4$~M$_\odot$), comparable to the high-mass end of our CGC range. In the supersonic regime, these clouds subsequently shatter into distinct clumps, closely matching the behaviour we observe in our simulations.

Our results strongly hint at a radiative turbulent mixing layer origin, as proposed in cloud-crushing studies \citep[e.g.][]{Gronke2018} Firstly, the cool cloud population vanishes in the absence of radiative cooling \citep{Ward2024}. In addition, \citet{Ward2024} and \citet{Ward2025} find the cooling time is particularly short in a mixed outflowing gas phase comprising a combination of wind fluid and ISM material.
Furthermore, we find the internal density profile of cool cloudlets to be approximately log-normal, consistent with \citet{BandaBarragan2020, BandaBarragan2021}. We also find a cumulative mass distribution $dN/dM \propto M^{-2}$, in agreement with the turbulent mixing layer prediction \citep{Gronke2022, Warren2024} (see Section~\ref{subsec:Impact-Initial-ISM} and Figure~\ref{fig:M-slope-time}).

\subsection{Limitations and future work}\label{subsec:limits}

Our simulations should be regarded as controlled experiments, isolating the effect of varying $\lagn$ and initial ISM structure on outflow properties. 
Their main strength is the extremely high resolution achieved in the outflow and, in particular, the cool phase. Our resolution of $M_{\rm target} \sim 1 \ \msun$ is not affordable in cosmological simulations, where typically $M_{\rm target} \gtrsim 10^3 \ \msun$.
Another strength of our simulations is their use of a fully physical AGN wind solution, reproducing the expected dynamics, energetics and multi-phase structure in exquisite detail \citep{Costa2020}.
Our simulations were designed to probe the impact of AGN winds on clumpy media, going beyond analytic understanding focussing on spherical wind solutions \citep{King2003, Faucher2012, Zubovas2012}. The setup is, however, clearly simplified, neglecting e.g. disc rotation, velocity dispersion and gravity, and further developments are needed to fully test:

\textit{Magnetic fields:} The presence of magnetic fields could significantly alter the dynamics of the simulated galaxy \citep{Avillez2005, Wang2009, vandeVoort2021}. Magnetic fields introduce additional pressure components that could confine gas bubbles and impede the mixing of hot and cold gas phases, both critical for the formation and stability of CGCs. 
Results from \cite{Avillez2005} showed the magnetic component dominates only for temperatures below $200$ K, in the range between $200$ K and $10^{5.5}$ K, the ram pressure dominates the dynamics. Other-hand, \cite{vandeVoort2021} (see their Figure 2) found when magnetic fields are included, the distribution of pressures becomes more strongly peaked, leading to a smoother circumgalactic medium. 

\textit{Self-Gravity}. Self-gravity can affect the size and properties of CGCs. Self-gravity promotes collapse, potentially leading to more pronounced fragmentation \citep{Arroyo2022, Mandal2024}, which can result in smaller, denser CGCs. Recent hydrodynamic simulations by \cite{Mandal2024} investigate turbulent, star-forming clouds interacting with high-pressure AGN-driven outflows and examine the role of self-gravity. Their findings suggest that small, fragmented clouds can become gravitationally bound, significantly extending their survival times. Future simulations should integrate both magnetohydrodynamic and self-gravity and test how these may affect the density - power relation discussed in this paper.

\textit{Optical Emission Lines}. Given the temperature range of our CGCs, we would expect them to emit in optical lines such as [O~{\sc iii}], [S~{\sc ii}], and H$\beta$. Estimating these emission lines could provide a more direct comparison with observational studies. However, any current estimations would not be self-consistent, as they do not include metal line cooling effects and realistic ionisation states, which will require radiative transfer. We plan to focus on this analysis in a future work.

\textit{Resolution}. Characterising the intricate structure of CGCs, such as those observed in our simulations, presents a significant computational challenge, particularly when aiming to capture fragmentation on sub-parsec scales. 
Our current resolution corresponds to a minimum mass scale of approximately $1 \, \msun$, which is sufficient to resolve the fragmentation of large ISM clumps with $M \gtrsim 10^4 \, \msun$. For instance, a clump with initial mass $10^4 \, \msun$, size $167$ pc (see Table~\ref{tab:simulations-params}), and average density $100$ cm$^{-3}$ --as adopted in our initial conditions (see \textit{W24})-- yields a clump diameter-to-resolution ratio of $r_{\rm cl} / D_{\rm cell} \approx 167 \text{ pc} / 0.78 \text{ pc} \approx 215$. This value is comparable to the resolution criteria outlined in \cite{Klein1994, Nakamura2006}, $r_{\rm cl} / D_{\rm cell}\geq 100$.
For fragmented clouds such as those seen in our simulations (see Figure~\ref{fig:maps}), the relevant characteristic scale is the cooling length, given by $c_s t_{\rm cool} \sim 0.1 \, (n / \text{cm}^{-3})^{-1}$ pc \citep{McCourt2018, Gronke2018, Gronke2020}. If this scale is not resolved, cloudlets may mix with the background at the grid scale. For gas with $n_{\rm H} = 10^2 \, \text{cm}^{-3}$, this implies a resolution requirement of $\lesssim 10^{-4}$ pc, and for denser gas with $n_{\rm H} = 10^4 \, \text{cm}^{-3}$, $\lesssim 10^{-6}$ pc; corresponding to a mass resolution of $m \sim 3 \times 10^{-5} \, \msun \, (n / \text{cm}^{-3})^{-2}$. Such resolution is far beyond what is currently feasible in galaxy-scale simulations.
Nonetheless, even though we do not resolve the full internal structure of the cloudlets, our simulations still allow us to observe their formation and destruction driven by AGN wind activity. For a more complete understanding of their evolution and properties, we plan to conduct dedicated high-resolution cloud-crushing experiments to investigate the wind–ISM interaction in greater detail.

\section{Conclusions} \label{sec:summary}

We have analysed hydrodynamic simulations of multiphase, energy-driven outflows emerging through the interaction between a small-scale AGN-driven wind and a clumpy, multiphase ISM. We have used the hydrodynamic simulations presented in \citet{Ward2024} along with new simulations that enhance resolution in outflowing, cool gas, as detailed in Section \ref{subsec:high-res-sims}. We have extracted and characterized a sub-population of cool gas clouds (CGCs), with temperatures $T < 2 \times 10^4$ K, within the simulated outflows (Section \ref{subsec:find-clouds}), and investigated their properties in relation to ISM porosity (Section \ref{subsec:Impact-Initial-ISM}) and wind power (Section \ref{subsec:Impact-LAGN}).

Our findings are:

\begin{enumerate}
    \item \textit{A density - luminosity relation for cool outflows}: We find that stronger winds, launched by brighter AGN, are associated with smaller and denser cool outflowing clouds (see Figures \ref{fig:maps} and \ref{fig:L-hist}). Strong AGN winds compress these cool clouds effectively, such that cool gas within the AGN outflow becomes denser than in the ambient ISM. The density and size of cool outflowing clouds scale as $n_\mathrm{H} \approx 50 \left( \frac{\lagn}{10^{43} \text{ erg s}^{-1}} \right)^{1/2}$ cm$^{-3}$ and $R_{\rm CGC} \approx 5 \left( \frac{\lagn}{10^{43} \text{ erg s}^{-1}} \right)^{-1/6}$ pc, respectively.

    \item \textit{Dependence on initial ISM clump size}: The initial clump sizes of the ISM have only moderate impact on the properties of cool, outflowing clouds, as evidenced in Figure \ref{fig:k-hist-t41}. While wind power has a more significant effect, we observe that the simulation with smaller initial clumps, $\lambda = 40$pc, produce denser CGCs compared to simulations with $\lambda = 167$pc or $333$pc by up to a factor of two. Simulations with initial clumps sizes $\lambda = 167$pc and $333$pc yielded very similar results. 
    Over time, we find that the influence of the initial conditions on the outflowing cool gas markedly decreases. Although convergence was not reached within the simulated time-frame, as the original ISM gas mixed with the wind and cools, the mass, density, and velocity distributions of CGCs across all models evolve toward similar shapes. 

    \item \textit{Dependence on Injected Wind Velocity}: Variations in the injected wind velocity also affect CGC properties. Increasing the wind velocity leads to a proportional increase in injected energy, resulting in denser and smaller CGCs. However, the mass and velocity distributions of the CGCs remained unchanged as we explore different injection velocities. 
    
    \item \textit{Hot phase vs. cool phase}: We observe that different phases react differently to changes in AGN power. The density of the shocked wind ($T \gtrsim 10^9 \, \rm K$) scales linearly with $\lagn$. Conversely, the outflowing cool phase ($T < 2 \times 10^4$K), follows a relation of $n_{\rm cool} \propto L_{\rm AGN}^{1/2}$ (see Figures \ref{fig:L-hist} and \ref{fig:pressure-radius}).
    
    \item \textit{Observational outflow rate estimates}: Our findings demonstrate a connection between cool gas density and wind power. Observational studies typically estimate the AGN mass outflow rate based on observations of optical emission lines, often calculating $\dot{M}_{\rm OF} \propto n_e^{-1}$ (see Equation \ref{eq:mass-outflow}), assuming constant electron density. By incorporating our correction factor due to the $\lagn$ dependency, as detailed in Equation \eqref{eq:mof-correction}, the estimated outflow rates are significantly impacted, sometimes by one or two orders of magnitude, which flattens the relationship between $\dot{M}_{\rm OF}$ and $\lagn$ (see Figure \ref{fig:fiore-comparison}).

    \item \textit{Implications for estimating kinetic coupling efficiency}: The kinetic coupling efficiencies for the highest-density gas phase are 0.05\% for $\lagn = 10^{45}$ erg s$^{-1}$ and 0.0015\% for $\lagn = 10^{47}$ erg s$^{-1}$ (see Figure \ref{fig:mdot_ek}). Despite seemingly low, these values should not be interpreted as indicative of inefficient AGN feedback. Indeed, for these luminosities, the AGN wind substantially modifies the gas distribution within the host galaxy, highlighting the complexity of AGN feedback mechanisms and suggesting that high-density gas is not a reliable tracer of kinetic coupling efficiency.
\end{enumerate}

Based on the results, we propose a new observational test to the energy-driven outflow scenario. While the emissivity of the hot, tenuous shocked wind component driving these outflows is extremely low \citep[e.g.][]{Nims2015}, we can infer its presence \emph{indirectly} through the effect its pressure has on the properties of (observable) cool gas embedded in the outflow. Systematic observational probes of outflow density (e.g. multiple measurements of electron densities which are sensitive to a broad dynamic range) across a wide range of AGN luminosity should look for a scaling relation. Its presence and slope will both provide important constraints on the driving mechanism of AGN outflows, leading to a fuller understanding about how these influence galaxy evolution.

\section*{Acknowledgements}
We used \texttt{Python} \citep{python2007, python2011} to organise the data and to make all figures. In this work we used several packages as \texttt{pandas} \citep{pandas}, \texttt{NumPy} \citep{numpy}, \texttt{SciPy} \citep{scipy} and \texttt{Matplotlib} \citep{matplotlib}.

We acknowledge useful discussions with Max Gronke, Luke Holden, Darshan Kakkad, Zhiyuan Yao, Aura Obreja, David Rosario, Vincenzo Mainieri, Matas Tartėnas, and Stephane V. Werner. We gratefully acknowledge Dr. Dipanjan Mukherjee for encouraging feedback and helpful suggestions, which enhanced the quality of this paper. IA and CH acknowledge funding from an United Kingdom Research and Innovation grant (code: MR/V022830/1). IA acknowledges support from the Deutsche Forschungsgemeinschaft (DFG, German Research Foundation) as part of the DFG Research Unit FOR5195 – project number 443220636.

This work used the DiRAC Memory Intensive service (Cosma8) at Durham University, managed by the Institute for Computational Cosmology, and the DiRAC Data Intensive service (CSD3) at the University of Cambridge, managed by the University of Cambridge University Information Services on behalf of the STFC DiRAC HPC Facility (www.dirac.ac.uk).
The DiRAC service at Durham was funded by BEIS, UKRI and STFC capital funding, Durham University and STFC operations grants. The DiRAC component of CSD3 at Cambridge was funded by BEIS, UKRI and STFC capital funding and STFC operations grants. DiRAC is part of the UKRI Digital Research Infrastructure. The original simulations from \Ward used computing facilities from the Computational Center for Particle and Astrophysics (C2PAP), part of the ORIGINS Excellence Cluster. The ORIGINS cluster is funded by the Deutsche Forschungsgemeinschaft (DFG; German Research Foundation) under Germany’s Excellence Strategy: EXC-2094-390783311.

\section*{Data Availability}

The data underlying this article will be shared on reasonable request to the corresponding author.


\bibliographystyle{mnras}
\bibliography{refs} 



\appendix
\appendix

\section{Cloud Finder Algorithm Tests}\label{app:DBSCAN-tests}

In order to identify CGCs within our simulations, as detailed in Section \ref{subsec:find-clouds}, we employed the extesively used DBSCAN algorithm \citep{Ester1996, Khan2014}. This choice was made to effectively distinguish between clustered elements (representative of gas cells in our simulation) and spurious noise elements within the computational domain.

\subsection*{Testing Setup}
\begin{figure}
    \centering
    \includegraphics[width=\linewidth]{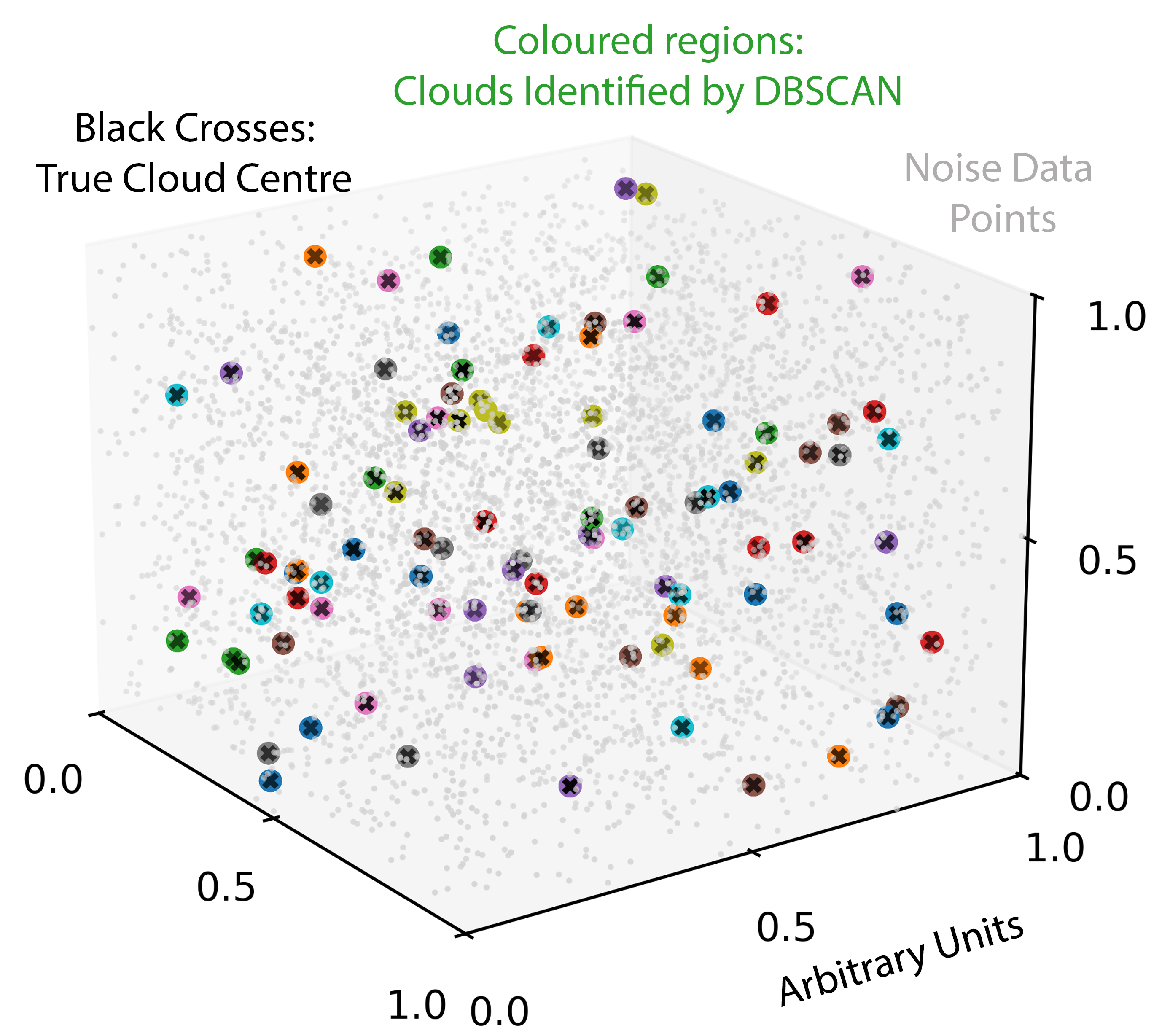}
    \caption{Test for our algorithm method. There are $100$ clouds within the cubic domain, with their centres marked by black crosses. The cloud elements are surrounded by grey noise elements. Using our algorithm, we found all clouds, and the cloud detection is shown as the coloured regions. This test demonstrates the efficiency of DBSCAN in finding clusters of elements based on their positions.}
    \label{fig:DBSCAN-test-100}
\end{figure}

For testing, we considered a cubic domain of unitary size, populated with arbitrarily placed elements. These elements represent the positions of gas cells in a typical application in our simulations. The domain contained a random distribution of clustered elements, simulating the distribution of gas clouds, along with a uniform distribution of random cells acting as noise elements. The distribution of these elements is shown in Figure \ref{fig:DBSCAN-test-100}, where the grey points represent the noise, and the black crosses mark the centres of the generated clouds. The clouds consist of $\approx 100$ elements within a radius of $10^{-3}$. In the test shown in Figure \ref{fig:DBSCAN-test-100}, there are $100$ clouds, comprising $10,000$ elements in total, plus $25,000$ noise elements.

DBSCAN was applied to this test setup to separate genuine clustered formations (the ``CGCs'') from the surrounding noise. The two main parameters our analysis depends on are the number of clustered cells $N_{\rm min}$ and the minimum distance to consider two cells neighbours $r_{\rm DB}$. The algorithm compares the positions of the elements and categorizes them into clustered packs or noise. The colored regions in Figure \ref{fig:DBSCAN-test-100} represent the detected clouds. For this plot, we used $N_{\rm min} = 10$ and $r_{\rm DB} = 10^{-3}$, and we detected precisely $100\%$ of the clouds.

\begin{figure}
    \centering
    \includegraphics[width=\linewidth]{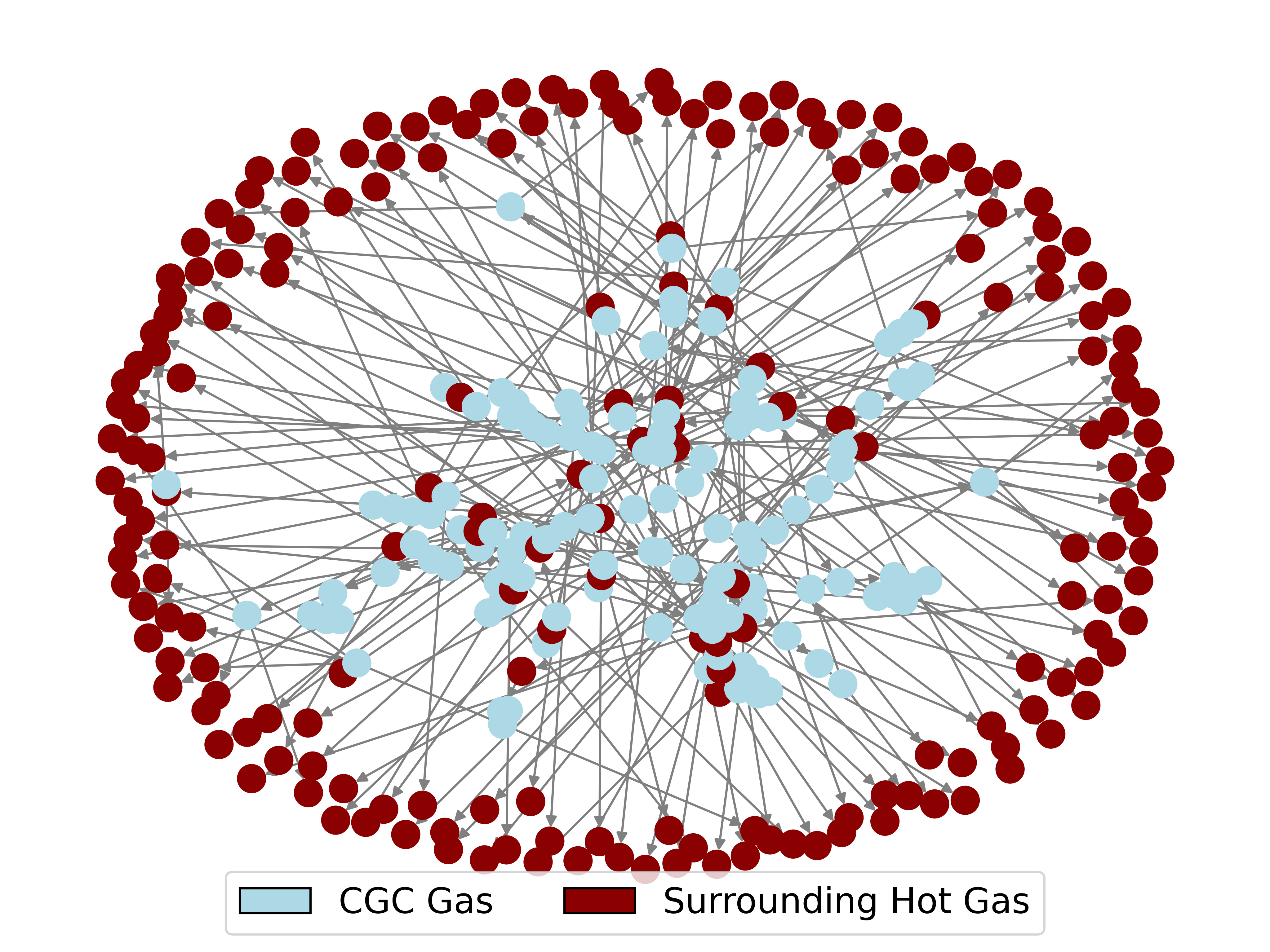}
    \caption{A neighbour tree for a random CGC. Blue circles represent the CGC members, and red circles are the surrounding hot gas cells. This visualization confirms that the cool gas cells are physically connected, forming a cohesive structure, while being surrounded by the hot gas.}
    \label{fig:CGC-tree}
\end{figure}

Another test for our method is presented in Figure \ref{fig:CGC-tree}. This plot visualizes the spatial connectivity within a single, randomly selected CGC and its surrounding environment. We tracked the neighbour cells for each cell within the CGC to understand how they interconnect. In the Figure, blue circles represent the cool gas cells of the CGC, while red circles depict the surrounding hot gas cells. The diagram clearly illustrates that the group of blue circles (cool gas) are interconnected, forming a coherent cloud, and are distinctly encapsulated by the red circles (hot gas) --there are some hot cells in the central region of the plot, this happens because the CGC has an irregular 3D structure, and \texttt{AREPO} uses a dynamic mesh.

\subsection*{Exploring Parameters}

\begin{figure}
    \centering
    \includegraphics[width=\linewidth]{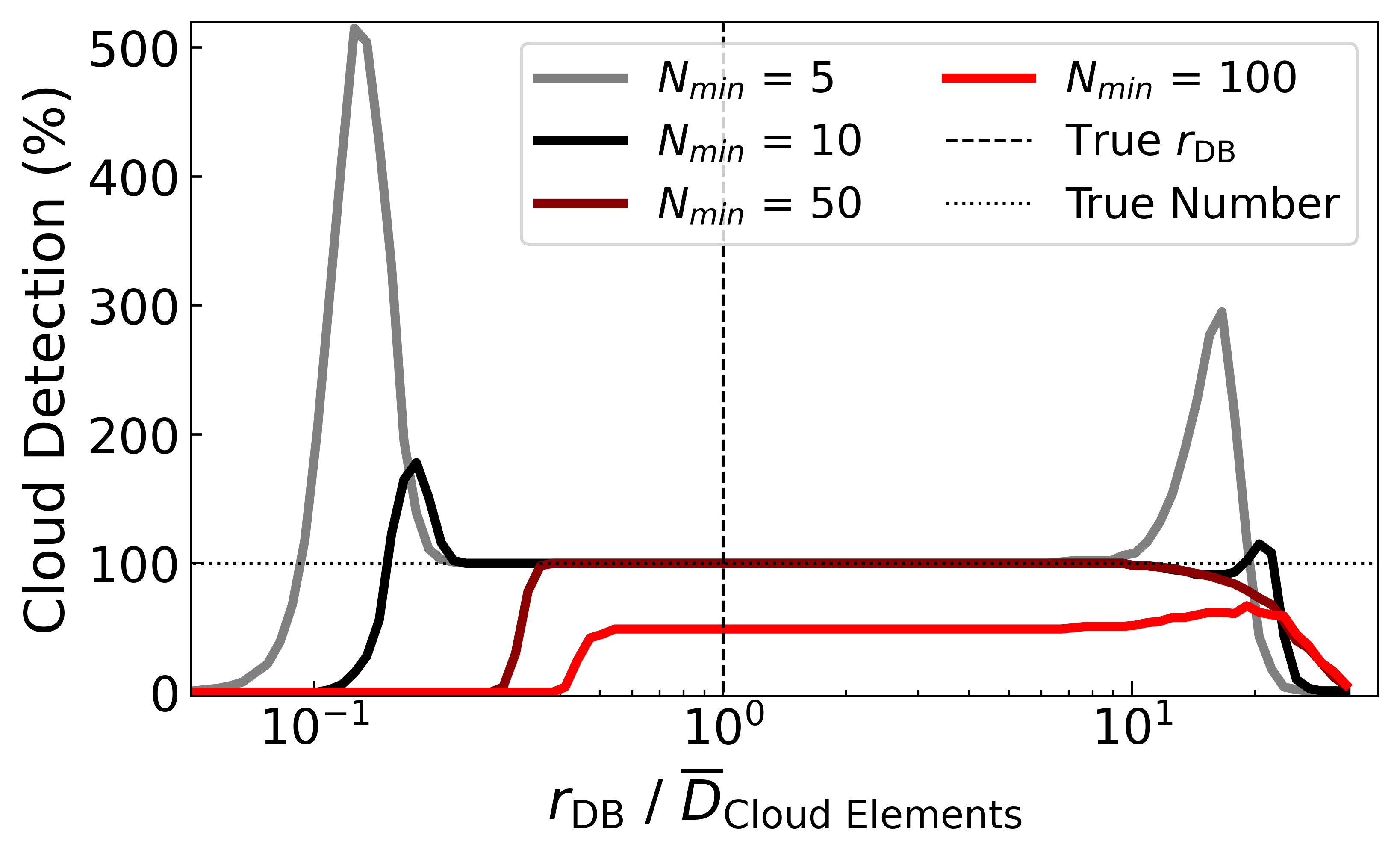}
    \caption{Cloud detection rate for the same setup as in Figure \ref{fig:DBSCAN-test-100}, varying $r_{\rm DB}/\overline{D}$ and $N_{\rm min}$ values, where $\overline{D}$ is the known average distance between elements within a cloud. The colors represent different $N_{\rm min}$ values. An optimal detection rate of $100\%$ is achieved for $r_{\rm DB}/\overline{D} \sim 0.5-10$ when $N_{\rm min}$ is set below the minimum number of elements known to be in any cloud. This figure illustrates the range of ideal values for the two parameters of our cloud finder algorithm, highlighting its efficacy under varying conditions.}
    \label{fig:DBSCAN-tests-rN}
\end{figure}

The detection accuracy of our cloud finder algorithm using DBSCAN is contingent upon the choices of $N_{\rm min}$ and $r_{\rm DB}$. We varied these parameters to assess their influence on detection efficacy. In our controlled test, the average distance between elements within the same cloud is known to be $10^{-3}$, and the average number of elements per cloud ranges between 80 and 120. Figure \ref{fig:DBSCAN-tests-rN} displays the number of detected clouds for the same setup as in Figure \ref{fig:DBSCAN-test-100}, while varying $N_{\rm min}$ and $r_{\rm DB}$. Here, $r_{\rm DB}$ is normalized by the average distance between cloud elements, a parameter known from the onset of this test but typically unknown in real applications. In the context of our CGCs, this value relates to the average cell size $\overline{D}$. This testing demonstrates that:

\begin{itemize}
    \item The optimal value of $r_{\rm DB}$ aligns with the structure's scale size and remains stable within the range $0.5\overline{D}$ to $10\overline{D}$.
    \item The ideal $N_{\rm min}$ is close to the minimum number of elements needed to form a cloud. Setting $N_{\rm min} = 50$ yields a well-behaved curve, whereas lower values cause over-prediction at small and large $r_{\rm DB}/\overline{D}$ ratios. Setting $N_{\rm min} = 100$ results in a well-behaved curve but restricts cloud detection to about 50\%, as half of our clouds contain fewer than 100 elements.
    \item Small values of $r_{\rm DB} \lesssim 0.2\overline{D}$ and lower ratios of $N_{\rm min}/N_{\rm true}$ lead to over-detection, counting a single structure multiple times.
    \item Conversely, higher values of $r_{\rm DB} \lesssim 10\overline{D}$ and smaller ratios of $N_{\rm min}/N_{\rm true}$ also lead to over-detection, capturing noise elements as clustered elements.
    \item Both over-prediction effects can be mitigated by increasing $N_{\rm min}$, which determines the minimum number of elements required for cloud classification, however, high-values of $N_{\rm min}$ will lead to under-detection.
\end{itemize}

DBSCAN's application in these tests proved highly effective at identifying CGCs and distinguishing them from spurious noise elements. This success in a controlled test environment supports the use of DBSCAN in our simulation data. For our analyses, we selected $N_{\rm min} = 100$, equivalent to a mass cut-off for CGCs of about $100 \ \msun$. This parameter was chosen to exclude numerical noise and remain above the resolution limit of approximately $2 \ \msun$. We also set $r_{\rm DB} = 4\overline{D}$, where $\overline{D}$ is the average size of a cool outflowing cell, equating to the average distance between CGC elements (see Section \ref{subsec:find-clouds}). This chosen $r_{\rm DB} = 4\overline{D}$ falls within the desirable range depicted in Figure \ref{fig:DBSCAN-tests-rN}.

\section{Acceleration Timescale}\label{app:acceleration-timescales}

In Section \ref{subsec:temporal-evolution}, we presented the velocity histograms for our CGCs and observed a velocity cut at $v \approx 300$km s$^{-1}$. None of the simulations identified CGCs with velocities $\gtrsim 500$km s$^{-1}$. Looking at the timescale for accelerating CGCs given by \citep{Gronke2018}:

\begin{equation}
    t_{\rm acc} = \chi^{1/2} t_{cc} = (2 \times 10^{-4} \text{ Myr}) \chi \frac{R_2}{v_4} 
    \label{eq:accelerating-timescale}
\end{equation}

\noindent where $t_{cc}$ is the cloud crushing timescale \citep{McCourt2018, Gronke2018} defined as:

\begin{equation}
    t_{\rm cc} = \chi^{1/2} \frac{R_{\rm CGC}}{v_{\rm wind}}
\label{eq:tcc-definition}
\end{equation}

Assuming the CGC density as given in equation \eqref{eq:lagn_x_n} and the wind density following the relation from equation 12 in \cite{Costa2020}, we derive:

\begin{equation}
    \chi = \frac{400 L_{45}^{1/2}}{31 L_{45} (D/1\text{ pc})^{-2} v_4^{-2}} = 1.3 \times 10^5 L_{45}^{-1/2}D_{100}^2
    \label{eq:chi-value}
\end{equation}

\noindent where $D_{100} = (D/100 \text{ pc})$ represents the distance of the cloud from the galaxy centre. We used $\beta = v_w/c = 0.1$ in \cite{Costa2014} formula to match the velocity of the injected wind at BOLA. Substituting equation \eqref{eq:chi-value} into equation \eqref{eq:accelerating-timescale} and using the $v_4 = 10^4$km s$^{-1}$ and $R_{\rm CGC} = 2L_{45}^{-1/6}$pc, which means $R_2 = L_{45}^{-1/6}$., we obtain:

\begin{equation}
    \begin{split}
        t_{\rm acc} = \left((2 \times 10^{-4} \text{ Myr}) \frac{R_2}{v_4}  \right) \left( 1.3 \times 10^5 L_{45}^{-1/2}D_{100}^2 \right)  \\
        = (26 \text{ Myr}) L_{45}^{-2/3}D_{100}^2v_4^{-1}
    \end{split} 
    \label{eq:accelerating-timescale-values}
\end{equation}

Most CGCs form at $D_{100} \geq 1$ (see Figure \ref{fig:vD-hist}). Consequently, the acceleration timescales for $\lagn = 10^{43}$ erg s$^{-1}$ and $\lagn = 10^{47}$ erg s$^{-1}$ and $v_4 =1$ are, respectively, $560$ Myr and $1.2$ Myr, both exceeding the duration of our simulations. This can explain the lack of high-velocity CGCs, accelerate these structures takes a very long time.

\bsp	
\label{lastpage}
\end{document}